\documentclass[aps,11pt,preprintnumbers,nofootinbib,floatfix]{revtex4}

\usepackage{graphicx}
\usepackage{epsfig}
\usepackage{dcolumn}
\usepackage{subfigure}
\usepackage{multirow}
\usepackage{amsmath}
\usepackage{amssymb}

\def\be{\begin{equation}}
\def\ee{\end{equation}}
\def\bea{\begin{eqnarray}}
\def\eea{\end{eqnarray}}

\begin{document}

\title{Implications for the hierarchy problem, inflation and geodesic motion from fiber fabric of spacetime}


\author{Cao H. Nam}
\email{nam.caohoang@phenikaa-uni.edu.vn}  
\affiliation{Phenikaa Institute for Advanced Study and Faculty of Fundamental Sciences, Phenikaa University, Yen Nghia, Ha Dong, Hanoi 12116, Vietnam}
\date{\today}

\begin{abstract}%

In this paper, we represent a resolution for the hierarchy problem where the inverse size of the extra dimension and the fundamental Planck scale would all be of the order of the TeV scale by proposing a fiber fabric of spacetime. The origin of the large hierarchy is essentially due to the $\cosh$ function which is physically originated from the dynamics of the horizontal metric in the vacuum of non-zero energy. In addition, the fiber fabric of spacetime allows us to resolve elegantly and naturally the problems of the chirality fermions and stabilizing potential for the size of the extra dimension, which are usually encountered in the higher dimensional theories. Then, we explore the inflation with the modulus of the extra dimension identified as the inflaton where our slow-roll inflationary model belongs to the E-model class with $n=1$. We calculate the main inflationary observables which are consistent with the present experiments. Finally, we study how the geodesic motion of neutral test particles gets modified from the extension of spactime. We compute the radius of the photon sphere, the innermost stable circular orbit, the perihelion shift, the light bending angle, and the observables of the strong gravitational lensing and the retrolensing phenomenon. By comparing the predicted values with the experimental observations, we determine the constraints on the fiber fabric of spacetime.

\end{abstract}

\maketitle

\section{Introduction}
General Relativity (GR) is one of the very successful theories of modern physics, besides the Standard Model (SM) of elementary particles, which describes many gravitational phenomena from the solar system scale upto the cosmological one to a high precision. In GR, spacetime is a four dimensional Lorentz manifold whose geometry is defined in terms of the metric field. Although the predictions of GR are well compatible to the observations, there are strong indications that the spacetime picture described by GR should be broken down at (very) short distances or high energies. For example, GR predicts the presence of Big Bang/black hole singularity, which was pointed by Hawking and Penrose that the formation of this spacetime singularity is inevitable under the certain conditions, or the quantization of gravity in GR is perturbatively nonrenormalizable. In addition, GR cannot explain inflation, dark matter, and dark energy relating to the late accelerating expansion of universe.

The problems of GR thus motivate alternative theories of gravity or the extensions of four dimensional spacetime. On the other hand, it is possible that spacetime fundamentally possesses the extra dimensions and certain hidden topological/geometric structures corresponding to new fields which become significant as probing into the short distances or the high energies. The generalization of GR to five dimensions were first proposed by Kaluza \cite{Kaluza1921} and Klein \cite{Klein1926} in attempting to unify the electromagnetic interaction with GR. However, the Kaluza-Klein (KK) theory encountered many problems such as all zero modes corresponding to the observed particles to be electrically neutral and hence it is inconsistent with the experimental observations. At the end of the 20th century, the extra dimensions have attracted enormous attention because they could provide a resolution for the problems of particle physics \cite{Dienes1998,Antoniadis1990,Arkani-Hamed2000}.

In particular, introducing the extra dimensions could resolve the hierarchy problem. In the scenario of the large extra dimensions proposed by Arkani-Hamed, Dimopoulos and Dvali (well-known as the ADD model), the inverse size of the extra dimensions is expressed as $R^{-1}=2\pi M_*(M_*/M_{\text{Pl}})^{2/n}$ where $M_*$ is the fundamental Planck scale, $M_{\text{Pl}}$ is the four dimensional or observed Planck scale $\sim{O}(10^{18})$ $\text{GeV}$, and $n$ is the number of the extra dimensions \cite{Arkani-Hamed1998,Antoniadis1998}. It is interesting that the fundamental Planck scale $M_*$ which is relevant to the quantum gravity scale would be lowered to the scales much lower than the observed Planck scale such as the TeV scale and thus the effects of quantum gravity might be accessible at the future colliding experiments. With the fundamental Planck scale $M_*$ of the order of a few TeV, the large hierarchy between the quantum gravity scale and the electroweak scale $\sim\mathcal{O}(1)$ TeV would no longer appear and thus the hierarchy problem is resolved in the scenario of the large extra dimensions. But, here the inverse size of the extra dimensions is so small compared to the fundamental Planck scale. This means that it leads to a new hierarchy problem.\footnote{An alternative for the hierarchy problem is based on the warped extra dimensions \cite{Randall1999} where the fundamental Planck scale and the inverse size of the extra dimensions would be of the order of the observed Planck scale.} As we will indicate in the present work, this new hierarchy problem can be resolved with a fiber fabric of five dimensional spacetime where the inverse size of the fifth dimension and the fundamental Planck scale would all be of the order of the TeV scale. There is another resolution for the hierarchy problem at which $M_*$ and $R^{-1}$ are also considred to be of the order of the TeV scale \cite{Kaloper2000}. In this work, the authors considered the compact hyperbolic extra dimensions where the volume of the extra space is exponentially amplified as, $\sim e^a$, with the topological parameter $a$ properly chosen without being fixed by a physical mechanism.

In the scenario of the large extra dimensions, all of the SM fields must live on a $3$-brane because the SM has been well tested up to the TeV scale. However, if the inverse size of the extra dimensions is of the order of the TeV scale or above, the SM fields can propagate in the higher dimensional spacetime \cite{Nath1999,Masip1999,Casalbuoni1999,Carone2000,Appelquist2001}. The problem arising here is that the higher dimensional theories give rise the spectrum of vector-like fermions, while the fermions in the SM are in fact chiral, which means that the left-handed and right-handed fermions transform in a different way under the gauge symmetry group. The problem of the chirality fermions could be overcome by introducing the orbifolds with singularities such as $S^1/Z_2$ and imposing the boundary conditions consistent with the fields \cite{Csaki2004}. But, it requires many technical assumptions which make the higher dimensional theories less nature. Therefore, it is worth investigating alternatives which are more elegant and nature.

There is another important problem of the higher dimensional theories based on the generalization of GR to the higher dimensions: what is the mechanism to fix the size of the extra dimensions or the mass of the radion field? In such higher dimensional theories, there is no the potential for the radion field because its potential which respects to the higher dimensional general covariance is constructed by contracting the indices of the metric to those of its dual, which would lead to a cosmological constant. As a result, the radion field is massless and thus it is inconsistent with the observations.

The inflation scenario provides an explanation for the horizon, flatness, and magnetic monopole problems \cite{Starobinsky1980,Guth1981,Linde1983,Albrecht1982}. In addition, it provides a mechanism to generate the primordial density fluctuations in the early universe which are the origin of the formation of the large scale structure as seen in the today universe. In order to solve the above problems, the inflation scenario predicts that the early universe undergoes a short period of accelerating expansion, which is simply driven by a scalar field called the inflaton. What is the physical origin of the inflaton or more explicitly whether it comes from particle physics or alternative theories of gravity? This still remains an open question and an active field of research. There have been the investigations which explore the possibility of the radion field as a natural candidate for the inflaton \cite{Cline2000,Kolb2003,Sundrum2010,Trudeau2012,Fukazawa2013}. Here, the potential of the radion field is effectively generated by coupling to the matter fields or quantum effects and breaks explicitly the higher dimensional general covariance. The potential of the radion field in the present work is different from the previous works as follows: it exists even at the tree level and respects to the higher dimensional general covariance.

The different theories alternative to GR can be tested by investigating the geodesic motion of the massless and massive test particles around the gravitational bodies. Particularly, the observations in the regime of the strongly gravitational field provide the promising opportunities to probe the viability range of GR and the hidden structures of spacetime. There exists the photon sphere, around the supermassive black holes, which is a set of the circular orbits of the photon. The photon sphere has been explored in the different black hole solutions \cite{Bahamonde2019,GuoMiao2020,Boonserm2020,Turimov2020,XXZeng2020,GuoLi2020}. In addition, the observations are obtained from the investigation of star motion around the supermassive black holes, such as the innermost stable circular orbit (ISCO) or the perihelion shift. ISCO of the massive particles is an important property for accretion disc formed around a black hole and it is the subject of special interest because its observation can derive the estimation on the parameters of the black hole as well as impose the constraints on alternative theories of gravity. The perihelion shift and the light bending angle in the solar system are much small compared to $2\pi$, however, one expects them to be large in the case of the strong gravitational bodies.

The present paper is structured as follows. In Sec. \ref{FS}, we propose a five dimensional extension of spacetime with a fiber fabric which is nontrivial in general. Then, we represent the intrinsic properties of spacetime and explore its implications for the hierarchy problem, the stabilizing potential for the size of the extra dimension, and the problem of the chirality fermions. In Sec. \ref{infla}, we investigate the inflation with the modulus of the extra space identified as the inflaton. In Sec. \ref{Geoeq}, we determine the geodesic equation and study its astrophysical implications. We consider the photon sphere, the radius of ISCO, the oscillating point of the near-circular orbits, the perihelion shift, the bending angle of the light in the weak field limit, and the strong gravitational lensing corresponding to the gravitational deflection of light in the strong field limit. Finally, we conclude in Sec. \ref{conclu}.

Throughout the present paper, we use the natural system of units ($\hbar=c=1$) and the signature of the metric as $(+,-,-,-,\cdots,-)$.

\section{\label{FS}Fundamental spacetime}

In the extra dimension theories which are the generalization of GR to $4+d$ dimensions with $d\geq1$ to be the number of extra dimensions and respecting to $4+d$ dimensional general covariance, one starts with $4+d$ dimensional (Lorentz) spacetime manifold $\mathcal{M}_{4+d}$ which can be decomposed as 
\begin{eqnarray}
\mathcal{M}_{4+d}=\bigcup_{i}U^{(4+d)}_i,\label{SPM-GGR}
\end{eqnarray}
where every $U^{(4+d)}_i$ is a portion (a local region) of the manifold $\mathcal{M}_{4+d}$ and looks like $\mathbb{R}^{4+d}$. Next, we assume that the ground state of the system which is obtained by the compactification has the topology $\mathcal{M}_4\times K_d$ (where $\mathcal{M}_4$ refers to the standard four dimensional spacetime manifold and $K_d$ is a $d$ dimensional compact extra-space), rather than $\mathcal{M}_{4+d}$ \cite{Salam1982}. But, as we mentioned in introduction, the extension in this way usually leads to
\begin{itemize}

\item[(i)] What is the stabilizing potential for the size of the extra dimension or the mass of the radion field.

\item[(ii)] The problem of the chirality fermions if they propagate in the bulk.

\end{itemize}

In order to introduce the higher dimensional extension of spacetime and overcome these problems in an elegant and  natural way, we do not consider spacetime manifold as decomposed by (\ref{SPM-GGR}), but with the following decomposition
\begin{eqnarray}
\mathcal{M}_{4+d}=\bigcup_{i}\left[U^{(4)}_i\times G\right],
\end{eqnarray}
where every $U^{(4)}_i\times G$ is a portion of $\mathcal{M}_{4+d}$, $U^{(4)}_i$ looks like $\mathbb{R}^4$, and $G$ is a Lie group manifold of $d$ dimensions. Such a spacetime manifold is realized as as a fiber bundle with the fibers to be identified as $G$. For the mathematical details, we refer the reader to Ref. \cite{Nakahara2003}. The fiber bundle structure allows us to solve the problems (ii), whereas the fibers to be the Lie group allows us to solve the problem\footnote{This will be clearly realized when one considers the Lie group to be non-Abelian.} (i).

In the present paper, we explore various aspects of the fiber fabric of spacetime and its phenomenological implications for the case of five dimensions, for simplicity. Our five dimensional spacetime $\mathcal{M}_5$ is a fiber bundle manifold whose typical fiber is the Lie group $U(1)$.

\subsection{General coordinate transformation}

By definition, one local region of spacetime $\mathcal{M}_5$ looks like $\mathbb{R}^4\times U(1)$ and hence it allows us to define the local coordinates for an arbitrary point in $\mathcal{M}_5$ as $\left(x^\mu,e^{i\theta}\right)$ where $\{x^\mu\}\in\mathbb{R}^4$ and $e^{i\theta}\in U(1)$ with $\theta$ to be dimensionless and real. When two local coordinate systems $U_i\times U(1)$ and $U_j\times U(1)$ are assigned to a same point in spacetime $\mathcal{M}_5$, we have the general coordinate transformation as
\begin{eqnarray}
x^\mu&\longrightarrow&x'^\mu=x'^\mu(x),\nonumber\\
e^{i\theta}&\longrightarrow&e^{i\theta'}=h(x)e^{i\theta},\ \
\textrm{or}\ \
\theta\longrightarrow\theta'=\theta+\alpha(x).\label{ba-gct}
\end{eqnarray}
This is the general coordinate transformation under which the theory constructed is required to be covariant.

We can consider the fundamental domain for the fiber coordinate $\theta$ as $\theta\in[-\pi,\pi]$ where the point with the fiber coordinates $\theta=-\pi$ and $\theta=\pi$ is clearly as the same.\footnote{One can consider other fundamental domains for the fiber coordinate $\theta$, such as $\theta\in[0,2\pi]$, without changing the physical results.} Accordingly, a field $\Phi(x,\theta)$ on spacetime $\mathcal{M}_5$ should satisfy the boundary condition.
\begin{eqnarray}
\Phi(x,-\pi)=\Phi(x,\pi).\label{boundcond2}
\end{eqnarray}

\subsection{Intrinsic properties}

Let $T_p\mathcal{M}_5$ be the tangent space to the spacetime manifold $\mathcal{M}_5$ at an arbitrary point $p\in\mathcal{M}_5$. There is a one dimensional subspace of $T_p\mathcal{M}_5$ which consists of all tangent vectors passing through $p\in\mathcal{M}_5$ which are tangent to the fiber or run along the fiber, called the vertical tangent subspace at $p\in\mathcal{M}_5$ and denoted as $V_p\mathcal{M}_5$. In this sense, $T_p\mathcal{M}_5$ can always be decomposed into a direct sum of the vertical tangent subspace and its complement without reference to the local coordinate system as follows
\begin{equation}
T_p\mathcal{M}_5=H_p\mathcal{M}_5\oplus V_p\mathcal{M}_5,
\end{equation}
where $H_p\mathcal{M}_5$ has dimension $4$ and is called the horizontal tangent subspace. Clearly, $H_p\mathcal{M}_5$ consists of all tangent
vectors (at $p\in\mathcal{M}_5$) which are transversal to the fiber. Note that, the directions to be transversal to the fibres are identified as the directions in the usually four dimensional spacetime. 

It is easy to see that the horizontal tangent subspace $H_pM_5$ is not spanned by $\{\partial_\mu\}$ because the ordinary partial derivative $\partial_\mu$ does not transform as a four dimensional vector under the general coordinate transformation (\ref{ba-gct}). But, it is spanned by the following basis
\begin{eqnarray}
\Big\{\partial_\mu-g_{_X}X_\mu\partial_\theta\Big\}\equiv\left\{\hat{\partial}_\mu\right\},
\end{eqnarray}
where the gauge field $X_\mu$ transforms under the general coordinate transformation (\ref{ba-gct}) as
\begin{eqnarray}
X_\mu\longrightarrow
X'_\mu&=&\frac{\partial x^\nu}{\partial x'^\mu}\left(X_\nu-\frac{1}{g_{_X}}\partial_\nu\alpha(x)\right).\label{B-gautrn}
\end{eqnarray}
and $g_{_X}$ is the corresponding gauge coupling. It is important to note here that the presence of the gauge field $X_\mu$ does not have the origin from the metric because the line element has not been introduced so far, but it arises from the non-triviality of the fiber fabric of spacetime. Whereas, the vertical tangent subspace $V_p\mathcal{M}_5$ is spanned by the basis vector $\partial_\theta$ which transforms as a one dimensional vector under the general coordinate transformation (\ref{ba-gct}).

The space dual to the tangent space $T_p\mathcal{M}_5$ is called the cotangent space and denoted as $T^*_p\mathcal{M}_5$. Because of the intrinsic decomposition of $T_p\mathcal{M}_5$ as analyzed above, the cotangent space $T^*_p\mathcal{M}_5$ can always decomposed in a direct sum as follows
\begin{eqnarray}
T^*_p\mathcal{M}_5=H^*_p\mathcal{M}_5\oplus V^*_p\mathcal{M}_5,
\end{eqnarray}
where $H^*_p\mathcal{M}_5$ and $V^*_p\mathcal{M}_5$ are horizontal and vertical cotangent subspaces which are dual to $H_p\mathcal{M}_5$ and $V_p\mathcal{M}_5$, respectively. The basis which span $H^*_p\mathcal{M}_5$ and $V^*_p\mathcal{M}_5$ are $\{dx^\mu\}$ and $\{d\theta+g_{_X}X_\mu dx^\mu\}$ which transform as four dimensional and one dimensional covectors (one-forms) under the general coordinate transformation (\ref{ba-gct}), respectively.

Now, in order to define the geometry of the spacetime manifold $\mathcal{M}_5$, one introduces a metric or an inner product $G$ on it. The intrinsic decomposition of any vector into the vertical and horizontal components without reference to the local coordinate system suggests an inner product $G$ on the spacetime manifold $\mathcal{M}_5$ as
\begin{equation}\label{bulk-mertic}
  G(V_1,V_2)=G_H(V_{1H},V_{2H})+G_V(V_{1V},V_{2V}),
\end{equation}
where $V_{1H}$ ($V_{2H})$ and $V_{1V}$ ($V_{2V}$) refer to the horizon and vertical components of the vector $V_1$ ($V_2$), respectively. In the above inner product, $G_H$ refers to the horizontal metric and it is a tensor field whose value at an arbitrary $p\in\mathcal{M}_5$ belongs to the space $H^*_p\mathcal{M}_5\otimes H^*_p\mathcal{M}_5$, which is generally given by \begin{equation}
  G_H=g_{\mu\nu}(x,\theta)dx^\mu dx^\nu.
\end{equation}
While, $G_V$ refers to the vertical metric and it is a tensor field whose value at an arbitrary $p\in\mathcal{M}_5$ belongs to the space $V^*_p\mathcal{M}_5\otimes V^*_p\mathcal{M}_5$, which is generally given as
\begin{equation}
G_V=-T^2(x,\theta)\frac{(d\theta+g_{_X}X_\mu
dx^\mu)^2}{\Lambda^2},\label{Vermetr-exp}
\end{equation}
where $\Lambda$ refers to a constant of the energy dimension. Note that, $\Lambda$ appears here because $d\theta+g_{_X}X_\mu dx^\mu$ is dimensionless but $(d\theta+g_{_X}X_\mu dx^\mu)/\Lambda$ has the dimension of length and thus defines the infinitesimal length on the fiber. One can check that, under the general coordinate transformation (\ref{ba-gct}), the metric fields $g_{\mu\nu}(x,\theta)$ and $T(x,\theta)$ transform as
\begin{eqnarray}
g_{\mu\nu}(x,\theta)&\longrightarrow& g'_{\mu\nu}(x',\theta')=\frac{\partial x^\rho}{\partial x'^\mu}\frac{\partial x^\lambda}{\partial x'^\nu}g_{\rho\lambda}(x,\theta),\nonumber\\
T(x,\theta)&\longrightarrow& T'(x',\theta')=T(x,\theta).\label{gmnT-gct}
\end{eqnarray}

In the frame $\Big\{\hat{\partial}_\mu,\partial_\theta\Big\}\equiv\left\{\partial_M\right\}$, the coefficients of the Christoffel connection and the Riemann curvature tensor take the form as follows
\begin{eqnarray}
\Gamma^P_{MN}&=&\frac{G^{PQ}}{2}\left(\partial_MG_{NQ}+\partial_NG_{MQ}-\partial_QG_{MN}\right)
+\frac{G^{PQ}}{2}\left(C^O_{QM}G_{ON}+C^O_{QN}G_{OM}\right)+\frac{C^P_{MN}}{2},\label{Chris-conn}\\
\mathcal{R}^O_{MPN}&=&\partial_N[\Gamma^O_{PM}]-\partial_P[\Gamma^O_{NM}]+
\Gamma^Q_{PM}\Gamma^O_{NQ}-\Gamma^Q_{NM}\Gamma^O_{PQ}+C^Q_{PN}\Gamma^O_{QM},\label{Riem-cur}
\end{eqnarray}
where $C^P_{MN}$ are the non-holonomic functions and determine the commutator of any two frame fields as
\begin{equation}
\left[\partial_M,\partial_N\right]=C^P_{MN}\partial_P,
\end{equation}
and
\begin{eqnarray}
  G_{MN}=\textrm{diag}\left(g_{\mu\nu}(x,\theta),-\frac{T(x,\theta)^2}{\Lambda^2}\right),\ \ G^{MN}=\textrm{diag}\left(g^{\mu\nu}(x,\theta),-\frac{\Lambda^2}{T(x,\theta)^2}\right).\label{bulk-metr}
\end{eqnarray}  

\subsection{Action}
An action which describes the dynamics of the fields $X_\mu$, $g_{\mu\nu}$, and $T$ can be written as a combination of the five dimensional Einstein-Hilbert action and the classical (tree level) potential for the field $T$ as
\begin{eqnarray}
S_{\text{spacetime}}&=&M^3_*\int d^4xd\theta\sqrt{-G}
\left[\frac{\mathcal{R}^{(5)}}{2}-f(T)\right]\nonumber\\
&=&\frac{M^3_*}{\Lambda}\int
d^4xd\theta\sqrt{-g}\left[\frac{T\hat{\mathcal{R}}}{2}+\frac{\Lambda^2}{8T}\left(\partial_\theta g^{\mu\nu}\partial_\theta g_{\mu\nu}+g^{\mu\nu}g^{\rho\lambda}\partial_\theta g_{\mu\nu}\partial_\theta g_{\rho\lambda}\right)\right.\nonumber\\
&&\left.-\frac{g^2_{_X}T^3}{8\Lambda^2}g^{\mu\rho}g^{\nu\lambda}X_{\mu\nu}X_{\rho\lambda}+\frac{3}{T}g^{\mu\nu}\left(\hat{\partial}_\mu T\right)\left(\hat{\partial}_\nu T\right)-V(T)\right],\label{EH-act}
\end{eqnarray}
where $\mathcal{R}^{(5)}$ is the scalar curvature of spacetime $\mathcal{M}_5$ which is explicitly expanded in the appendix, $\hat{\mathcal{R}}$ is denoted as
\begin{eqnarray}
\hat{\mathcal{R}}\equiv g^{\mu\nu}(\hat{\partial}_\nu\Gamma^\lambda_{\lambda\mu}-\hat{\partial}_\lambda\Gamma^\lambda_{\nu\mu}+\Gamma^\rho_{\lambda\mu}\Gamma^\lambda_{\nu\rho}-\Gamma^\rho_{\nu\mu}\Gamma^\lambda_{\lambda\rho}),
\end{eqnarray}
with $\Gamma^\rho_{\mu\nu}\equiv\frac{g^{\rho\lambda}}{2}(\hat{\partial}_\mu g_{\lambda\nu}+\hat{\partial}_\nu g_{\lambda\mu}-\hat{\partial}_\lambda g_{\mu\nu})$, $X_{\mu\nu}\equiv\partial_\mu X_\nu-\partial_\nu X_\mu$ is the field strength tensor of the gauge field $X_\mu$ which is related to the commutation of two horizontal basis vectors as
\begin{eqnarray}
[\hat{\partial}_\mu,\hat{\partial}_\nu]=-g_{_X}X_{\mu\nu}\partial_\theta,
\end{eqnarray} 
and $V(T)\equiv Tf(T)$ is the potential of the field $T$ at the tree level which can be constructed due to the fact that $T$ transforms as a scalar under the general coordinate transformation (\ref{ba-gct}) and is generally given as follows
\begin{eqnarray}
V(T)=\Lambda^2T\left[a_0+a_1T+a_2T^2+a_3T^3+a_4T^4+\cdots\right],\label{Tpot}
\end{eqnarray}
with $a_{i}$ to be the dimensionless constants, which follows a Taylor series expansion of $f(T)$. As seen below, when we normalize canonically $T$ as given in Eq. (\ref{Cannor-Tf}), only the first two terms in the potential $V(T)$ have the significant contribution as seen in Eq. (\ref{sta-pot}). Whereas, the higher order terms of the potential in terms of the canonically normalized field $T'$ would be highly suppressed by the further powers of $1/M_*$ and hence they are well-approximately ignored. This suggests that we can impose the mathematical conditions on the potential $V(T)$ as follows
\begin{eqnarray}
\frac{\partial^{(n)}V}{T'^{(n)}}\Big|_{T'=0}=0 \ \ \ \ \text{with} \ \ n>4,
\end{eqnarray}
which corresponds to the mathematical conditions on $f(T)$ as
\begin{eqnarray}
\frac{\partial^{(n)}\left(T'^2f\right)}{T'^{(n)}}\Big|_{T'=0}=0 \ \ \ \ \text{with} \ \ n>4.
\end{eqnarray}

Note that, from the action (\ref{EH-act}), we see that no kinetic term in the vertical direction, which is proportional to $G^{\theta\theta}\partial_\theta T\partial_\theta T$, is obtained from the scalar curvature $\mathcal{R}^{(5)}$. This suggests that without loss of generality we can consider the field $T$ independent on the coordinate $\theta$, i.e. $T=T(x)$, in the following calculations.

Before proceeding, let us clarify the following point: the field $T$ is just part of the vertical metric tensor given in Eq. (\ref{Vermetr-exp}); however, this field appears as an independent field in the action $S_{\text{spacetime}}$. In the expression of the vertical metric tensor, $T^2$ is its component in the basis $\left\{(d\theta+g_{_X}X_\mu dx^\mu\right)^2\}$. When changing the basis of the vertical metric tensor, its component or the field $T$ transforms in the law given by Eq. (\ref{gmnT-gct}). In this way, the vertical metric tensor may be represented by the field $T$ equipped with the transformation law (\ref{gmnT-gct}). (In general, any tensor may be represented by its components equipped with a transformation law which encodes the information of how its components respond to the change of the basis). This means that, in order for the terms related to the field $T$ in the action to be independent on the choice of the basis, they must be invariant under the transformation law (\ref{gmnT-gct}).

The field $T$ can be canonically normalized with the change of the field variable as
\begin{eqnarray}
T'=\left(\frac{24M^3_*}{\Lambda}T\right)^{1/2}.\label{Cannor-Tf}
\end{eqnarray}
Then, the corresponding potential is 
\begin{eqnarray}
V(T')\simeq\mu^2_{T'}T'^2+\lambda_{T'}T'^4,\label{sta-pot}
\end{eqnarray}
where 
\begin{eqnarray}
\mu^2_{T'}&\equiv&\frac{a_0\Lambda^2}{24},\nonumber\\
\lambda_{T'}&\equiv&\frac{a_1\Lambda^3}{576M^3_*}
\end{eqnarray}
and the higher order terms in terms of $T'$ in $V(T')$ are highly suppressed by the further power of $1/M_*$ and hence they are well-approximately ignored. The nonzero vacuum expectation value (VEV) of the field $T'$ corresponding to the minimum of the potential $V(T')$ is obtained as
\begin{eqnarray}
\langle T'\rangle\simeq\sqrt{-\frac{\mu^2_{T'}}{2\lambda_{T'}}}.
\end{eqnarray}
As a result, the size $R$ of the fifth dimension is physically determined as 
\begin{eqnarray}
R\equiv\frac{\langle T\rangle}{\Lambda}=\frac{\langle T'\rangle^2}{24M^3_*}.
\end{eqnarray} 

We study the dynamics of the horizontal metric $g_{\mu\nu}$ in the vacuum $T'=\langle T'\rangle$ at which the vacuum energy is non-zero in general. Hence, first let us write the equation of motion for the horizontal metric $g_{\mu\nu}$ in the presence of vacuum energy described by a cosmological constant $\Lambda_{\text{cc}}$ as
\begin{eqnarray}
\bar{R}_{\mu\nu}-\frac{1}{2}g_{\mu\nu}\bar{R}+\frac{1}{4R^2}\left[g_{\mu\rho}g_{\nu\lambda}\partial^2_\theta g^{\rho\lambda}-\partial^2_\theta g_{\mu\nu}+2g^{\rho\lambda}\partial_\theta g_{\rho\lambda}\partial_\theta g_{\mu\nu}\right.&&\nonumber\\
\left.+2\partial_\theta\left(g_{\mu\nu}g^{\rho\lambda}\partial_\theta g_{\rho\lambda}\right)-\frac{1}{2}g_{\mu\nu}\left(\partial_\theta g^{\mu\nu}\partial_\theta g_{\mu\nu}+g^{\mu\nu}g^{\rho\lambda}\partial_\theta g_{\mu\nu}\partial_\theta g_{\rho\lambda}\right)\right]+\Lambda_{\text{cc}}g_{\mu\nu}&=&0,
\end{eqnarray}
where the expressions of $\bar{R}_{\mu\nu}$ and $\bar{R}$ are same to the conventional four dimensional Ricci and scalar curvatures, respectively, but written in terms of $g_{\mu\nu}(x,\theta)$, and the cosmological constant $\Lambda_{\text{cc}}$ is written more explicitly as 
\begin{eqnarray}
\Lambda_{\text{cc}}=-24\lambda_{T'}\langle T'\rangle^2+\cdots,
\end{eqnarray}
with the first term coming from the value of the potential $V(T')$ at the minimum $T'=\langle T'\rangle$ and the ellipse referring to the contribution expected from other sources such as the potential of the scalar fields and zero-point energy caused by the quantum fluctuations. Note that, we have assumed that the contributions coming from the fluctuation around $\langle T'\rangle$, the gauge field $X_\mu$ and other fields are small compared to the vacuum energy and since they are negligible. 

We solve this equation by separating the variables as $g_{\mu\nu}(x,\theta)=\chi(\theta)\bar{g}_{\mu\nu}(x)$ where $\bar{g}_{\mu\nu}(x)$ is identified as the effectively four dimensional metric. From the boundary condition (\ref{boundcond2}) on $g_{\mu\nu}(x,\theta)$, we find the corresponding boundary condition on $\chi(\theta)$ as
\begin{eqnarray}
\chi(-\pi)=\chi(\pi).\label{bouncond-chi}
\end{eqnarray} 
Then, we obtain the equations for $\bar{g}_{\mu\nu}(x)$ and $\chi(\theta)$ as
\begin{eqnarray}
\mathcal{R}^{(4)}_{\mu\nu}-\frac{1}{2}\bar{g}_{\mu\nu}\mathcal{R}^{(4)}+\frac{k}{4R^2}\bar{g}_{\mu\nu}&=&0,\label{effEinsEq}\\
6\chi''(\theta)+4\frac{\chi'(\theta)^2}{\chi(\theta)}-\frac{4\Lambda_{\text{cc}}}{\langle G^{\theta\theta}\rangle}\chi(\theta)&=&k,\label{chiEq}
\end{eqnarray}
where $\mathcal{R}^{(4)}_{\mu\nu}$ and $\mathcal{R}^{(4)}$ are the conventional four dimensional Ricci and scalar curvatures corresponding to the effectively four dimensional metric $\bar{g}_{\mu\nu}(x)$, respectively, and $k$ is a constant.
We have comments as follows: 
\begin{itemize}
\item[1.] The equation for $\chi(\theta)$ is a nonlinear differential equation and hence there is no a general solution as the linear combination of partial solutions.
\item[2.] The vertical kinetic part of the horizontal metric leads to a contribution to the vacuum energy in the four dimensional effective theory.
\item[3.] Eq. (\ref{effEinsEq}) is Einstein field equations in the four dimensional effective theory with a cosmological constant $\Lambda^{(4)}_{\text{cc}}$ which is given by
\begin{eqnarray}
\Lambda^{(4)}_{\text{cc}}=\frac{k}{4R^2}.\label{effCC}
\end{eqnarray} 
If $\Lambda^{(4)}_{\text{cc}}$ is identified as the observed cosmological constant, by using the corresponding experimental value $\Lambda^{(4)}_{\text{cc}}\approx5.06\times10^{-84}$ $\text{GeV}^2$ \cite{Tanabashi2018}, we determine $k$ in terms of the inverse size of the fifth dimension as
\begin{eqnarray}
k&\simeq&2\times10^{-83}\times\frac{\text{GeV}^2}{R^{-2}}.
\end{eqnarray}
This relation suggests that $k$ is positive and extremely small even $R^{-1}$ is of the order of the electroweak scale. Due to this fact in what follows we shall consider $k\geq0$.
\end{itemize}

We now find the solutions for Eq. (\ref{chiEq}) which satisfy the boundary condition (\ref{bouncond-chi}) for the following three cases:\\
\emph{Case 1.} $\Lambda_{\text{cc}}<0$: To obtain the generally analytical solution with $k\neq0$ is a difficult task. However, one can find the perturbative solution which has the following expansion
\begin{eqnarray}
\chi(\theta)=\chi^{(0)}(\theta)+\sum_{n=1}k^n\chi^{(n)}(\theta),\label{pertsol}
\end{eqnarray}
where $\chi^{(0)}(\theta)$ is the solution at the leading order corresponding to $k=0$ (corresponding to the exact vanishing of the cosmological constant in the effectively four dimensional theory) and $\chi^{(n)}(\theta)$ are the solutions at the subleading orders in $k$. Due to the fact that $k$ is extremely small as indicated above, the solution can be well-approximated by $\chi^{(0)}(\theta)$.

The solution with $k=0$ is easily obtained as follows
\begin{eqnarray}
\chi(\theta)=C_1\cosh^{3/5}\left[\frac{\sqrt{10}}{3}\kappa_1(\theta+C_2)\right],\label{negLacc1}
\end{eqnarray}
where $\kappa_1\equiv\sqrt{-\Lambda_{\text{cc}}/R^{-2}}$ and $C_{1,2}$ are the constants. Note that, $C_1$ can be set to be unity by the fact that $C_1$ can be absorbed into the effective metric $\bar{g}_{\mu\nu}(x)$. While, the boundary condition (\ref{bouncond-chi}) suggests $C_2=0$. 

It is interesting that there is a particularly and non-trivially analytical solution as
\begin{eqnarray}
\chi(\theta)=\frac{5k}{28\kappa^2_1}\left[\cosh\left(\sqrt{\frac{2}{5}}\kappa_1\theta\right)-1\right].\label{negLacc2}
\end{eqnarray}
This solution clearly respects to the boundary condition (\ref{bouncond-chi}).\\
\emph{Case 2.} $\Lambda_{\text{cc}}=0$: We can find the analytical solution is
\begin{eqnarray}
\chi(\theta)= 
\left\{%
\begin{array}{ll}
    \left[-\chi^{5/3}{_2F_1}\left(\frac{1}{2},\frac{5}{7};\frac{12}{7};-\frac{k\chi^{7/3}}{7}\right)\right]^{(-1)}(\theta), & \hbox{$-\pi\leq\theta<0$;} \\
    \left[\chi^{5/3}{_2F_1}\left(\frac{1}{2},\frac{5}{7};\frac{12}{7};-\frac{k\chi^{7/3}}{7}\right)\right]^{(-1)}(\theta), & \hbox{$0\leq\theta\leq\pi$,} \\
\end{array}%
\right.\label{solchi-c22}
\end{eqnarray}
where $_2F_1(a,b;c;z)$ is the hypergeometric function and $[\cdots]^{(-1)}$ is denoted the inverse function which means
\begin{eqnarray}
\theta=\mp\chi^{5/3}{_2F_1}\left(\frac{1}{2},\frac{5}{7};\frac{12}{7};-\frac{k\chi^{7/3}}{7}\right).
\end{eqnarray}
Of course, the solution in this case satisfies the boundary condition (\ref{bouncond-chi}).\\
\emph{Case 3.} $\Lambda_{\text{cc}}>0$: In analogy to the case of $\Lambda_{\text{cc}}<0$, the analytical solution at the leading order in $k$ is given as
\begin{eqnarray}
\chi(\theta)=\left\{%
\begin{array}{ll}
    \cos^{3/5}\left(\frac{\sqrt{10}}{3}\kappa_2\theta\right), & \hbox{$-\frac{\pi}{2}+2m\pi\leq\frac{\sqrt{10}}{3}\kappa_2\theta<\frac{\pi}{2}+2m\pi$;} \\
    \cos^{3/5}\left(\frac{\sqrt{10}}{3}\kappa_2\theta-\pi\right), & \hbox{$\frac{\pi}{2}+2m\pi\leq\frac{\sqrt{10}}{3}\kappa_2\theta<\frac{3\pi}{2}+2m\pi$,} \\
\end{array}%
\right.\label{Ca1k0}
\end{eqnarray}
where $\kappa_2\equiv\sqrt{\Lambda_{\text{cc}}/R^{-2}}$ and $m\in\mathbb{Z}$. The boundary condition (\ref{bouncond-chi}) imposes on the possible values of $\kappa_2$ as
\begin{eqnarray}
\kappa_2=\frac{3}{\sqrt{10}} n,\ \ \ \ n=1,2,3,\cdots.\label{kap2val1}
\end{eqnarray}
In addition, we find a particularly and non-trivially analytical solution as
\begin{eqnarray}
\chi(\theta)=\frac{5k}{28\kappa^2_2}\left[1+\cos\left(\sqrt{\frac{2}{5}}\kappa_2\theta\right)\right],\label{Ca1kn0}
\end{eqnarray}
where the parameter $\kappa_2$ is not quantized by (\ref{kap2val1}), but by
\begin{eqnarray}
\kappa_2=\sqrt{\frac{5}{2}} n,\ \ \ \ n=1,2,3,\cdots.\label{kap2val2}
\end{eqnarray}
Here, we note that for two cases with $k=0$ and $k\neq0$, there exists the trivial solution as
\begin{eqnarray}
\chi(\theta)=\text{constant},\label{trisol}
\end{eqnarray}
which obviously satisfies the boundary condition (\ref{bouncond-chi}) and can be set to be unity.

From the above analysis, we can rewrite the effective action from $S_{\text{spacetime}}$ as
\begin{eqnarray}
S^{\text{eff}}_{\text{spacetime}}&=&\int
d^4x\sqrt{-g}\left[\frac{\widetilde{T}^2}{24}\frac{\mathcal{R}^{(4)}}{2}-\frac{1}{4}X^{\mu\nu}X_{\mu\nu}\left(1+\cdots\right)+\frac{\lambda_{\widetilde{T},X}}{2}\widetilde{T}^2X^\mu X_\mu\right.\nonumber\\
&&\left.+\frac{1}{2}\partial_\mu\widetilde{T}\partial^\mu\widetilde{T}-\lambda_{\widetilde{T}}\left(\widetilde{T}^2-\langle \widetilde{T}\rangle^2\right)^2\right],\label{effSPact}
\end{eqnarray}
where we have redefined $T'$ such that the corresponding kinetic term is canonical as
\begin{eqnarray}
\widetilde{T}&=&T'\left[\int^\pi_{-\pi}d\theta\chi(\theta)\right]^{1/2},\nonumber\\
\lambda_{\widetilde{T}}&=&\lambda_{T'}\left[\int^\pi_{-\pi}d\theta\chi^2(\theta)\right]\left[\int^\pi_{-\pi}d\theta\chi(\theta)\right]^{-2},
\end{eqnarray}
due to the integration on $\theta$, also in order to have the canonically normalized action for the gauge field
$X_\mu$ the gauge coupling $g_{_X}$ has been fixed as
\begin{eqnarray}
g_{_X}&=&\frac{R^{-1}}{M_{\text{Pl}}}\left[\frac{1}{\pi}\int^\pi_{-\pi}d\theta\chi(\theta)\right]^{1/2},
\end{eqnarray}
$\lambda_{T',X}$ is given by
\begin{eqnarray}
\lambda_{T',X}&=&\frac{R^{-2}}{4\pi M^2_{\text{Pl}}}\int^\pi_{-\pi}d\theta\left[\chi''(\theta)-\frac{\chi'(\theta)^2}{2\chi(\theta)}\right],
\end{eqnarray}
and the ellipse refers to the terms relating to the fluctuation field around the vacuum $\langle\widetilde{T}\rangle$. Note that, first in obtaining the effective action $S^{\text{eff}}_{\text{spacetime}}$, we have used the fact that $\chi(\theta)$ and $\chi'(\theta)$ are even and odd functions over the fundamental domain $[-\pi,\pi]$, respectively, which suggests 
\begin{eqnarray}
\int^\pi_{-\pi}d\theta\mathcal{F}[\chi(\theta)]\chi'(\theta)=0,
\end{eqnarray}
where $\mathcal{F}[\chi(\theta)]$ is an arbitrary function of $\chi(\theta)$. Second, we have considered the vanishing of the effective cosmological constant due to its near-zero value. This situation is suitable to the present observation of the cosmological constant as well as the inflation scenario because the inflation caused by the energy coming from the potential of the inflaton (which is identified as the field $\widetilde{T}$ in this work) stops when the inflaton rolls down to the global minimum of the potential at which its value is equal to or near zero. Third, the indexes appearing in the effective action $S^{\text{eff}}_{\text{spacetime}}$ are contracted together by the effectively four dimensional metric $\bar{g}_{\mu\nu}(x)$.

We observe from the effective action $S^{\text{eff}}_{\text{spacetime}}$ that the gauge field $X_\mu$ can become massive in the effective theory with the mass squared given as
\begin{eqnarray}
m^2_X&=&\lambda_{\widetilde{T},X}\langle\widetilde{T}\rangle^2\nonumber\\
&=&\left\{%
\begin{array}{ll}
    \frac{27}{80\pi}\bar{\kappa}_1 \left\{\frac{56\sinh(\bar{\kappa}_1\pi)}{\cosh^{\frac{2}{5}}(\bar{\kappa}_1\pi)}-15\cosh^{\frac{8}{5}}(\bar{\kappa}_1\pi)\Re\left[i\,_2F_1\left(\frac{1}{2},\frac{4}{5};\frac{9}{5};\cosh^2(\bar{\kappa}_1\pi)\right)\right]\right\}\frac{1}{R^2}, &  \hbox{(I);} \\
    \frac{9\times10^{-83}}{7\pi\kappa_1}\left[\sqrt{\frac{5}{2}}\sinh\left(\sqrt{\frac{2}{5}}\kappa_1\pi\right)-\kappa_1\pi\right]\times1\ \ \text{GeV}^2, & \hbox{(II);} \\
    \frac{35R^{-2}}{\pi\chi^{2/3}}\left[\chi\,_2F_1\left(\frac{1}{7},\frac{1}{2};\frac{8}{7};-\frac{1}{7} k \chi^{7/3}\right)-\frac{9k \chi^{7/3}}{280}\, _2F_1\left(\frac{1}{2},\frac{5}{7};\frac{12}{7};-\frac{1}{7} k \chi^{7/3}\right)\right.\\
    \left.+\frac{21+8\chi}{28}\left(1+\frac{1}{7} k \chi^{7/3}\right)^{1/2}\right]\Big|^{\chi=\chi_0}_{\chi=0}, & \hbox{(III);} \\
    -\frac{27n^2R^{-2}}{4\pi\cos^{2/5}\theta}\Big|_{\theta\rightarrow\pi/2}, & \hbox{(IV);} \\
    - \frac{9\times10^{-83}}{7}\times1\ \ \text{GeV}^2, & \hbox{(V).} \\
    0, & \hbox{(VI).} \\
\end{array}%
\right.\label{Xgaufimas}
\end{eqnarray}
where $\bar{\kappa}_1\equiv\frac{\sqrt{10}}{3}\kappa_1$, $\chi_0$ is the solution of the following equation
\begin{eqnarray}
\chi^{5/3}_0{_2F_1}\left(\frac{1}{2},\frac{5}{7};\frac{12}{7};-\frac{k\chi^{7/3}_0}{7}\right)=\pi,
\end{eqnarray}
the cases (I), (II), (III), (IV), (V), and (VI) correspond to the solutions given in Eqs. (\ref{negLacc1}), (\ref{negLacc2}), (\ref{solchi-c22}), (\ref{Ca1k0}), (\ref{Ca1kn0}), and (\ref{trisol}), respectively. Note here that, we have used the identification between the observed Planck scale $M_{\text{Pl}}$ and ${\langle\widetilde{T}\rangle}$ as $M_{\text{Pl}}=\frac{\langle\widetilde{T}\rangle}{2\sqrt{6}}$ which allows us to fix the VEV of the field $\widetilde{T}$ as
\begin{eqnarray}
\langle\widetilde{T}\rangle=2\sqrt{6}M_{\text{Pl}}.\label{VEVT-4DMp}
\end{eqnarray}
We make a few comments on the mass of the gauge field obtained from various solutions of $\chi(\theta)$. It is easily to see that the mass-square of the gauge field $X_\mu$ corresponding to the cases (III) and (IV) gets the infinitely negative values. This suggests the inconsistency of the theory and thus these cases should be excluded. In the case (V), the mass-square of the gauge field $X_\mu$ in the effective theory is negative but its value is extremely small and hence it can get other corrections to become positive. However, in this case first the corresponding gauge coupling is $g_{_X}=\sqrt{\frac{20}{7n^2}}\times\frac{10^{-41}\ \ \text{GeV}}{M_{\text{Pl}}}\lesssim7\times10^{-60}$ which is near zero and hence it almost decouples the effective theory. Second, because $k$ is extremely small it would lead to the very large hierarchy between the five dimensional Planck scale $M_*$ and the inverse size of the extra dimension in order to obtain the observed Planck scale $M_{\text{Pl}}$. This implies the solution (\ref{Ca1kn0}) less interesting.

\subsection{Resolving the hierarchy problem}
Now let us indicate that the present spacetime model would allow to resolve the hierarchy problem where there is no the large hierarchy between the fundamental Planck scale $M_*$, the electroweak scale, and the inverse size of the extra dimension. In this scenario, unlike the ADD model the inverse size of the extra dimension and the fundamental Planck scale would all be of the order of the TeV scale.

First, let us identify the observed Planck scale $M_{\text{Pl}}$ from the effective action $S^{\text{eff}}_{\text{spacetime}}$ as
\begin{eqnarray}
M^2_{\text{Pl}}&=&\frac{M^3_*}{R^{-1}}\int^\pi_{-\pi}d\theta\chi(\theta).\label{4DPlscal}
\end{eqnarray}
Then, for the solution given by Eq. (\ref{negLacc1}), the observed Planck scale $M_{\text{Pl}}$ is expressed in terms of the five dimensional Planck scale $M_*$, the inverse size of the extra dimension, and the parameter $\kappa_1$ as
\begin{eqnarray}
M^2_{\text{Pl}}&=&\frac{3}{4\kappa_1}\sqrt{\frac{5}{2}}\cosh^{\frac{8}{5}}\left(\frac{\sqrt{10}}{3}\kappa_1\pi\right)\Re\left[i\, _2F_1\left(\frac{1}{2},\frac{4}{5};\frac{9}{5};\cosh^2\left(\frac{\sqrt{10}}{3}\kappa_1\pi\right)\right)\right]\frac{M^3_*}{R^{-1}},
\end{eqnarray}
where $\Re$ is denoted the real part. Form this relation, we can address the hierarchy problem why the observed Planck scale $M_{\text{Pl}}\sim{O}(10^{18})$ $\text{GeV}$ is so much larger than the electroweak scale $\sim\mathcal{O}(1)$ TeV and the inverse size of the extra dimension even they are the same scale at the fundamental level. 

It is important to note that, because the SM has been tested up to the scale of a few TeV, the five dimensional Planck scale $M_*$ must be above $\mathcal{O}(1)$ TeV. Furthermore, the five dimensional Planck scale $M_*$ plays the role of the natural UV cutoff. This means that, from the definition which the parameter $\kappa_1$ is the ratio of the root-square modulus of the bulk cosmological constant to the inverse size of the extra dimension, the maximal value of $\kappa_1$ should be of the order of $M_*R$. 

\begin{figure}[t]
 \centering
\begin{tabular}{cc}
\includegraphics[width=0.6 \textwidth]{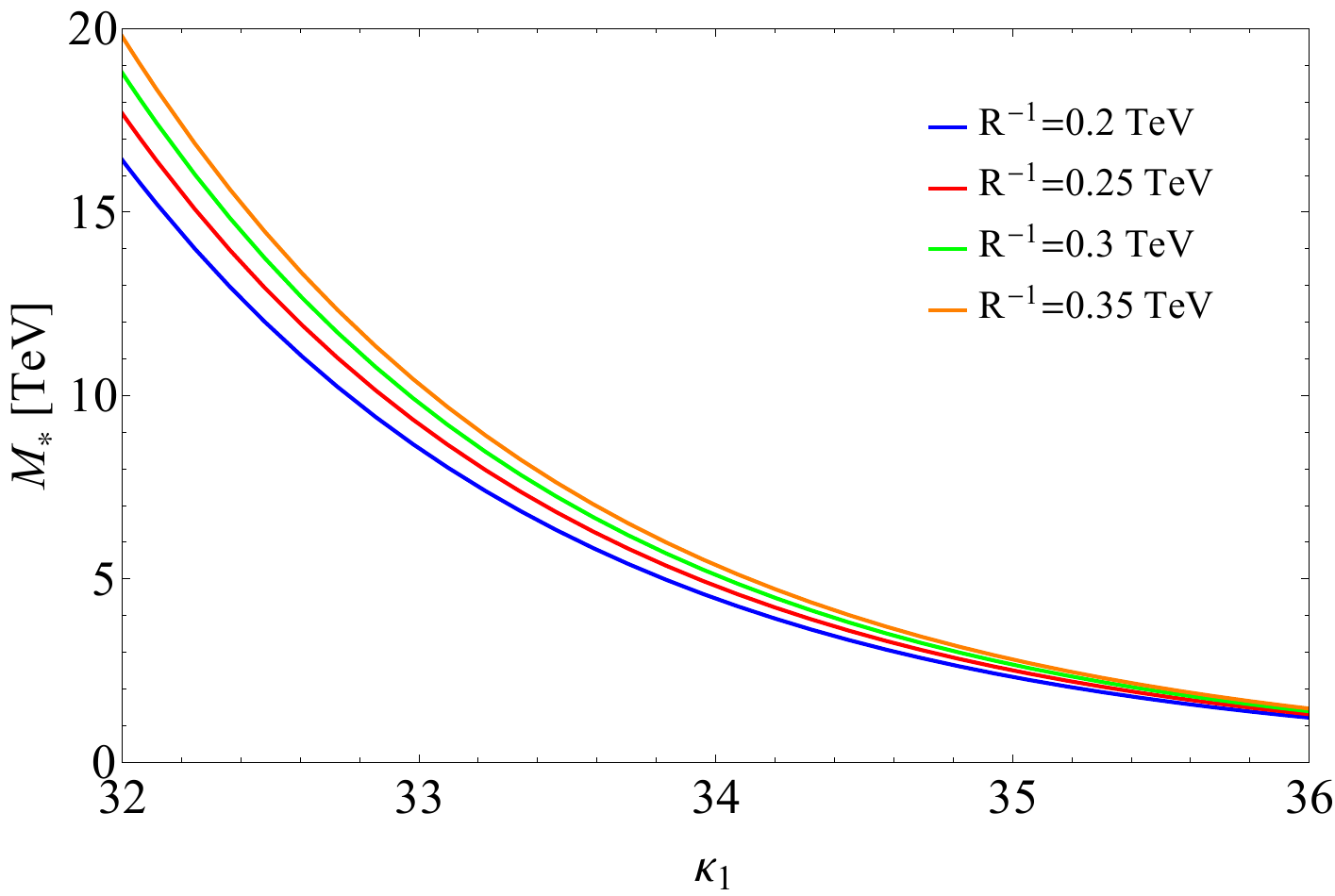}
\end{tabular}
  \caption{The dependence of the five dimensional Planck scale $M_*$ in the parameter $\kappa_1$ for various values of the inverse size of the extra dimension.}\label{M-kap1}
\end{figure}

If we take $R^{-1}\sim 0.3$ TeV and $M_*\sim6$ TeV, one finds $\kappa_1\sim33.8$ in order to obtain the observed Planck scale. For more explicitly, we show the behavior of the five dimensional Planck scale $M_*$ in terms of the parameter $\kappa_1$ for various values of the inverse size of the extra dimension in Fig. \ref{M-kap1}.
From this figure, we can see that both the five dimensional Planck scale $M_*$ and the inverse size of the extra dimension being of the order of the TeV scale require the parameter $\kappa_1\sim\mathcal{O}(33.5)$. Here, the five dimensional Planck scale $M_*$ is slightly larger than the inverse size of the extra dimension. It is easily to realize that the origin of the large hierarchy between the observed Planck scale $M_{\text{Pl}}$ and the electroweak scale is due to the function of $\cosh\left(\frac{\sqrt{10}}{3}\kappa_1\pi\right)$ which comes from the dynamics of the horizontal metric in the vacuum of non-zero energy and grows too fast in terms of increasing of the parameter $\kappa_1$. 

We note that we have assumed that the Higgs field lives on the $3$-brane which is localized around $\theta=0$ in order for $\chi(0)\sim1$. Since the physical electroweak scale does not get modified its order due to the normalization of the Higgs field.

\subsection{Fermions in bulk spacetime}
As we mentioned in the introduction, if the SM fermions propagate in the higher dimensional spacetime, the problem of the chirality fermions would arise because it gives rise the spectrum of vector-like fermions whereas the SM fermions are in fact chiral. In this subsection, we would like to indicate that with the fiber fabric of spacetime the conventional left-handed and right handed Weyl spinor fields can be in nature introduced even in five dimensional spacetime and hence the problem of the chirality fermions would disappear. Then, we shall determine the action which describes their propagation in spacetime $\mathcal{M}_5$ and obtain their KK spectrum.

One can write the horizontal metric $G_H$ at the local flat form by a $\text{GL}(4,\mathbb{R})$-rotation of the basis vectors as
\begin{eqnarray}
\hat{e}_\alpha&=&{e_\alpha}^\mu\hat{\partial}_\mu,\nonumber\\
\hat{e}^\alpha&=&{e^\alpha}_\mu dx^\mu,
\end{eqnarray}
where $\left\{\hat{e}_\alpha\right\}$ and $\left\{\hat{e}^\alpha\right\}$ are the new bases of the horizontal tangent and cotangent subspaces, respectively, ${e_\alpha}^\mu$ is called vierbeins and ${e^\alpha}_\mu$ is its inverse, i.e. ${e^\alpha}_\mu{e_\alpha}^\nu={\delta_\mu}^\nu$ and ${e^\alpha}_\mu{e_\beta}^\mu={\delta^\alpha}_\beta$. In the new basis, the horizontal metric becomes
\begin{eqnarray}
G_H=\eta_{\alpha\beta}\hat{e}^\alpha\hat{e}^\beta,\label{HM-lff}
\end{eqnarray}
where $\eta_{\mu\nu}=\text{diag}\left(1,-1,-1,-1\right)$ and ${e^\alpha}_\mu$ satisfies
\begin{eqnarray}
{e^\alpha}_\mu{e^\beta}_\nu\eta_{\alpha\beta}=g_{\mu\nu},
\end{eqnarray}
which implies ${e^\alpha}_\mu=\chi^{1/2}(\theta){\bar{e}^\alpha}_\mu(x)$ with ${\bar{e}^\alpha}_\mu(x)$ satisfying ${\bar{e}^\alpha}_\mu(x){\bar{e}^\beta}_\nu(x)\eta_{\alpha\beta}=\bar{g}_{\mu\nu}(x)$.

Of course, there are many alternative choices for $\left\{\hat{e}^\alpha\right\}$ which yield the same local flat form (\ref{HM-lff}) for the horizontal metric $G_H$, by the local rotation as
\begin{eqnarray}
\hat{e}^\alpha\longrightarrow\hat{e}'^{\alpha}={\Lambda^\alpha}_\beta\hat{e}^\beta,\label{Lolorsot}
\end{eqnarray}
where ${\Lambda^\alpha}_\beta$ satisfies ${\Lambda^\alpha}_\beta{\Lambda^\delta}_\gamma\eta_{\alpha\delta}=\eta_{\beta\gamma}$ and thus ${\Lambda^\alpha}_\beta\in SO(3,1)$. Accordingly, we introduce the (left-handed or right handed) Weyl spinor field $\Psi$ which transforms under the local rotation (\ref{Lolorsot}) as $\Psi\rightarrow\Psi'=\rho({\Lambda^\alpha}_\beta)\Psi$ where $\rho({\Lambda^\alpha}_\beta)$ is the (left-handed or right-handed) Weyl spinor representation of ${\Lambda^\alpha}_\beta$.

Now we introduce the action describing the dynamics of the Weyl spinor field $\Psi$ in spacetime $\mathcal{M}_5$ as
\begin{eqnarray}
S[\Psi]&=&\int d^4xd\theta\sqrt{-G}\left[
\bar{\Psi}i\gamma^\alpha {e_\alpha}^\mu\left(\hat{\partial}_\mu+\frac{i}{2}\omega^{\alpha\beta}_\mu\Sigma_{\alpha\beta}\right)\Psi+\frac{1}{2\Lambda}\left(\partial^\theta\bar{\Psi}\partial_\theta\Psi^C+\textrm{H.c.}\right)\right],\label{feract}
\end{eqnarray}
where $\gamma^\alpha$ are the Dirac matrices satisfying the Clifford algebra $\{\gamma^\alpha,\gamma^\beta\}=2\eta^{\alpha\beta}$, $\Sigma_{\alpha\beta}=\frac{i}{4}[\gamma_\alpha,\gamma_\beta]$ are the generators of the group $SO(3,1)$, $\omega^{\alpha\beta}_\mu$ is the spin connection expressed as
\begin{eqnarray}
\omega^{\alpha\beta}_\mu={e^\alpha}_\nu\left(\hat{\partial}_\mu e^{\beta\nu}+e^{\beta\lambda}\Gamma^\nu_{\mu\lambda}\right),
\end{eqnarray}
with $e^{\beta\nu}=\eta^{\beta\gamma}{e_\gamma}^\nu$, and $\Psi^C=C\Psi^*$ is the charge-conjugate field with $C=i\gamma^2\gamma^0$ and $\Psi^*$ to be the complex conjugate field. The first and second terms in the action (\ref{feract}) describe the dynamics of $\Psi$ along the horizontal and vertical directions in spacetime $\mathcal{M}_5$, respectively. It is important to note that if the field $\Psi$ transforms \textit{non-trivially} under a certain local gauge symmetry then the second term in the action (\ref{feract}) is not be invariant and hence forbidden. On the other hand, such a field has no the vertical kinetic term and thus it has no the KK counterparts.

In order to derive the KK spectrum of $\Psi$, let us consider its \textit{free} propagation corresponding to the propagation in the vacuum $\bar{g}_{\mu\nu}=\eta_{\mu\nu}$ [or ${e_\alpha}^\mu=\chi^{-1/2}(\theta){\delta_\alpha}^\mu$], $X_\mu=0$, and $\widetilde{T}=\langle\widetilde{T}\rangle$. The equation of motion for the field $\Psi$ reads
\begin{eqnarray}
\chi^{-1/2}(\theta)i\gamma^\mu\partial_\mu\Psi+\frac{R^{-2}}{\Lambda\chi^{2}(\theta)}\partial_\theta\left[\chi^2(\theta)\partial_\theta\Psi^C\right]=0.
\end{eqnarray}
We solve this equation by separating the variables as $\Psi(x,\theta)=\chi^{-3/4}(\theta)\psi(x)Y(\theta)$ where $Y(\theta)$ satisfies the boundary condition
\begin{eqnarray}
Y(-\pi)=Y(\pi),\label{Ybouncon}
\end{eqnarray}
which is obtained from Eq. (\ref{boundcond2}). Then, we find
\begin{eqnarray}
i\gamma^\mu\partial_\mu\psi(x)-m\psi(x)^C&=&0,\\
Y''(\theta)+\frac{\chi'(\theta)}{2\chi(\theta)}Y'(\theta)-\left[\frac{3\chi''(\theta)}{4\chi(\theta)}+\frac{3}{16}\left(\frac{\chi'(\theta)}{\chi(\theta)}\right)^2-a\chi^{-1/2}(\theta)\right]Y(\theta)&=&0,\label{fer-Yeq}
\end{eqnarray}
where $a$ is a dimensionless constant, $m=aR^{-2}/\Lambda$, and $\psi(x)^C=C\psi(x)^*$. Eq. (\ref{fer-Yeq}) can be rewritten in the form of Sturm-Liouville equation as
\begin{eqnarray}
\frac{d}{d\theta}\left[p(\theta)\frac{d\bar{Y}(\theta)}{d\theta}\right]+\left[q(\theta)+aw(\theta)\right]\bar{Y}(\theta)=0,
\end{eqnarray}
where $\bar{Y}(\theta)\equiv\chi^{-3/4}(\theta)Y(\theta)$, $p(\theta)=\chi^2(\theta)$, $q(\theta)=0$, and $w(\theta)=\chi^{3/2}(\theta)$.
Because $p(\theta)$ and $w(\theta)$ are the differentiable and positive functions over the fundamental domain $\theta\in[-\pi,\pi]$, all eigenvalues $\{a\equiv a_n\}$ are real and non-negative. In general, the solution of $\Psi(x,\theta)$ is given in terms of the KK expansion as
\begin{eqnarray}
\Psi(x,\theta)=\frac{\chi^{-3/4}(\theta)}{\sqrt{2\pi R}}\sum_n\psi_n(x)Y_n(\theta),
\end{eqnarray}
where $\{Y_n(\theta)\}$ corresponding to $\{a_n\}$ forms a complete set each of which satisfies Eq. (\ref{fer-Yeq}). By using the boundary condition (\ref{Ybouncon}) and Eq. (\ref{fer-Yeq}) for $Y_n(\theta)$, one finds
\begin{eqnarray}
(a_n-a_m)\int^\pi_{-\pi}d\theta Y_n(\theta)Y_m(\theta)=0.
\end{eqnarray}
Clearly, this relation suggests the orthogonality relation with respect to $a_n\neq a_m$.

With the nontrivial solution of $\chi(\theta)$, we can find the analytical solution for Eq. (\ref{fer-Yeq}) in the case of the small values of the parameters $\kappa_{1,2}$ (where we leave the hierarchy problem). In this situation, we are only interested in the nontrivial solution (\ref{negLacc1}), whereas the remaining nontrivial solutions (\ref{negLacc2}) and (\ref{Ca1kn0}) are less interesting because they lead to the very large hierarchy between the five dimensional Planck scale $M_*$ and the inverse size of the extra dimension due to the extreme smallness of $k$. Up to the order $\kappa^2_1$, the solution of Eq. (\ref{fer-Yeq}) is given as
\begin{eqnarray}
Y(\theta)&=&e^{-\frac{\kappa^2_1\theta^2}{12}\left(1+\sqrt{1+6a/\kappa^2_1}\right)}\left[CH\left(-\frac{4+\sqrt{1+6a/\kappa^2_1}-6a/\kappa^2_1}{2\sqrt{1+6a/\kappa^2_1}},\frac{\kappa_1\theta}{\sqrt{6}}(1+6a/\kappa^2_1)^{1/4}\right)\right.\nonumber\\
&&\left.+C'{_1F_1}\left(\frac{4+\sqrt{1+6a/\kappa^2_1}-6a/\kappa^2_1}{4\sqrt{1+6a/\kappa^2_1}};\frac{1}{2};\frac{\kappa^2_1\theta^2}{6}\sqrt{1+6a/\kappa^2_1}\right)\right],
\end{eqnarray}
where $C$ and $C'$ are the constants which can be fixed from the normalization condition, $H(a,z)$ and ${_1F_1}(a;b;z)$ are the Hermite polynomial and the confluent hypergeometric function, respectively. The relation (\ref{Ybouncon}) for $Y(\theta)$ imposes the possible values for $a$ as
\begin{eqnarray}
a\equiv a_n=\frac{\kappa^2_1}{12}\left[9+8n(1+2n)+(1+4n)\sqrt{21+8n(1+2n)}\right],
\end{eqnarray}
where $n=0$, $1$, $2$, $\cdots$ which are the non-negative integer numbers. Interestingly, the zero mode corresponding to $a_0=(9+\sqrt{21})\kappa^2_1/12$ has the non-zero Majorana mass 
as
\begin{eqnarray}
m_0=\frac{(9+\sqrt{21})\kappa^2_1}{12}\frac{R^{-2}}{\Lambda}.
\end{eqnarray} 

With the trivial solution of $\chi(\theta)$ given in Eq. (\ref{trisol}), it is straightforward to obtain the solution for Eq. (\ref{fer-Yeq}) as
\begin{eqnarray}
Y(\theta)=C'\cos(\sqrt{a}\theta)+C\sin(\sqrt{a}\theta),
\end{eqnarray}
where $a\equiv a_n=n^2$ with $n=0$, $1$, $2$, $\cdots$. The phenomenological implications for particle physics corresponding to this solution are explored in Refs. \cite{Nam2019a,Nam2020a,Nam2020b}. 
 
\section{\label{infla}Inflation}
In this section, we would like to indicate that our model provides a natural inflation scenario where the field $\widetilde{T}$ which is related to the geometry of the extra dimension plays the role of the inflaton. In this sense, the presence of inflation is unavoidable because the inflaton field is associated to the mathematical structure of spacetime.
 
We begin with writing the action for the effective metric field $\bar{g}_{\mu\nu}(x)$ and the field $\widetilde{T}$ as follows
\begin{eqnarray}
S^{\text{eff}}_{\text{spacetime}}\supset\int
d^4x\sqrt{-g}\left[\frac{\widetilde{T}^2}{24}\frac{\mathcal{R}^{(4)}}{2}+\frac{1}{2}\partial_\mu\widetilde{T}\partial^\mu\widetilde{T}-\lambda_{\widetilde{T}}\left(\widetilde{T}^2-\langle\widetilde{T}\rangle^2\right)^2\right],
\end{eqnarray}
which is in the so-called Jordan frame. The above action can be rewritten in the so-called Einstein frame via the conformal transformation as
\begin{eqnarray}
\hat{g}_{\mu\nu}=\Omega^2\bar{g}_{\mu\nu},\ \ \ \ \Omega^2=\frac{\widetilde{T}^2}{24M^2_{\text{Pl}}}.
\end{eqnarray}
This conformal transformation leads to a non-canonical kinetic term for the inflaton. Hence, we redefine $\widetilde{T}$ as
\begin{eqnarray}
\widetilde{T}=\langle\widetilde{T}\rangle e^{\frac{\chi}{\sqrt{30}M_{\text{Pl}}}},\label{chdef}
\end{eqnarray}
where $\chi$ is the field of the canonical kinetic term. Then, the potential for the field $\chi$ takes the form as
\begin{eqnarray}
U(\chi)&=&\frac{\lambda_{\widetilde{T}}\left(\widetilde{T}^2-\langle\widetilde{T}\rangle^2\right)^2}{\Omega^4}\nonumber\\
&=&576\lambda_{\widetilde{T}}M^4_{\text{Pl}}\left(1-e^{-\frac{2\chi}{\sqrt{30}M_{\text{Pl}}}}\right)^2.
\end{eqnarray}
This inflation potential $U(\chi)$ belongs to the slow-roll inflation class of E-model ($n=1$) \cite{Planck2018} with the parameter $\alpha^E_1=5$. The global minimum of the potential $U(\chi)$ is at $\langle\chi\rangle=0$ which corresponds to VEV of the original field $\widetilde{T}$ as $\langle\widetilde{T}\rangle$ as seen in Eq. (\ref{chdef}).

The slow-roll parameters are obtained from the inflation potential $U(\chi)$ as
\begin{eqnarray}
\epsilon&=&\frac{M^2_{\text{Pl}}}{2}\left[\frac{U'(\chi)}{U(\chi)}\right]^2,\nonumber\\
\eta&=&M^2_{\text{Pl}}\frac{U''(\chi)}{U(\chi)},\nonumber\\
\zeta^2&=&M^4_{\text{Pl}}\frac{U'''(\chi)U'(\chi)}{U^2(\chi)}.
\end{eqnarray}
These parameters are related to the observed quantities which are the spectral index $n_s$, the tensor-to-scalar ratio $r$, the running index $\frac{dn_s}{d\ln k}$, and the scalar perturbation amplitude $A_s$ as
\begin{eqnarray}
n_s&\simeq&1-6\epsilon+2\eta,\nonumber\\
r&\simeq&16\epsilon,\nonumber\\
\frac{dn_s}{d\ln k}&\simeq&16\epsilon\eta-24\epsilon^2-2\zeta^2,\nonumber\\
A_s&=&\frac{U}{24\pi^2M^2_{\text{Pl}}\epsilon},
\end{eqnarray}
which are evaluated at the horizon crossing value of the inflaton. Here, the observed values of the spectral and running indexes at $68\%$ confidence limit (CL) are $n_s=0.9649\pm0.0042$ and $\frac{dn_s}{d\ln k}=-0.0045\pm0.0067$, the scalar perturbation amplitude is measured as $A_s\approx2.2\times10^{-9}$, whereas the upper bound for the tensor-to-scalar ratio at $95\%$ CL is $r<0.063$, as reported by Planck 2018 for the pivot scale $0.05$ Mpc$^{-1}$ \cite{Planck2018}. The number of e-foldings is given as
\begin{eqnarray}
N=\int^{\chi_{\text{in}}}_{\chi_{\text{end}}}\frac{1}{M^2_{\text{Pl}}}\frac{U}{U'(\chi)}d\chi,\label{efold}
\end{eqnarray}
where $\chi_{\text{end}}$ is the value of the inflaton at which the inflation ends and $\chi_{\text{in}}$ is the horizon crossing value of the inflaton. By using $\epsilon\simeq1$ at the end of the inflation, we obtain $\chi_{\text{end}}\simeq1.14M_{\text{Pl}}$. 

It is straightforward to solve Eq. (\ref{efold}) to find $\chi_{\text{in}}$ and then the inflation observables. In Table \ref{Inftab}, we show our predictions for $\chi_{\text{in}}$, the inflation observables, and the coupling $\lambda_{\widetilde{T}}$ for various number of e-foldings.
\begin{table}[!htp]
\centering
\begin{tabular}{cccccc}
  \hline
  \hline
  $\ \ N \ \ $ & $\ \ \chi_{\text{in}}\ \ $ & $\ \ n_s\ \ $ & $\ \ r\ \ $ & $\ \ \frac{dn_s}{d\ln k} \ \ $ & $\ \ \lambda_{\widetilde{T}}\ \ $  \\
  \hline
  \hline
  $\ \  58\ \ $ & $ \ \ 8.14M_{\text{Pl}}\ \ $ & $\ \ 0.9681\ \ $ & $ \ \ 0.0124\ \  $ & $\ \ -5.30\times10^{-4}\ \ $ & $\ \ 7.79\times10^{-13}\ \ $\\
  \hline
  $59$ & $8.18M_{\text{Pl}}$ & $0.9686$ & $0.0120$ & $-5.12\times10^{-4}$ & $7.55\times10^{-13}$\\
  \hline
  $60$ & $8.22M_{\text{Pl}}$ & $0.9692$ & $0.0117$ & $-4.96\times10^{-4}$ & $7.32\times10^{-13}$\\
  \hline
  $61$ & $8.26M_{\text{Pl}}$ & $0.9696$ & $0.0114$ & $-4.80\times10^{-4}$ & $7.10\times10^{-13}$\\
  \hline
  $62$ & $8.29M_{\text{Pl}}$ & $0.9701$ & $0.0110$ & $-4.65\times10^{-4}$ & $6.89\times10^{-13}$\\
  \hline
  \hline
\end{tabular}
\caption{The inflationary predictions in our model for various number of e-foldings.}\label{Inftab}
\end{table}
From this table we find that our predictions are in good agreement with the Planck 2018 data. Compared to the inflationary models of the Starobinsky-type potential, our model predicts a larger tensor-to-scalar ratio which is around $\sim0.011$ and thus can be tested in the near future.

\begin{figure}[t]
 \centering
\begin{tabular}{cc}
\includegraphics[width=0.6 \textwidth]{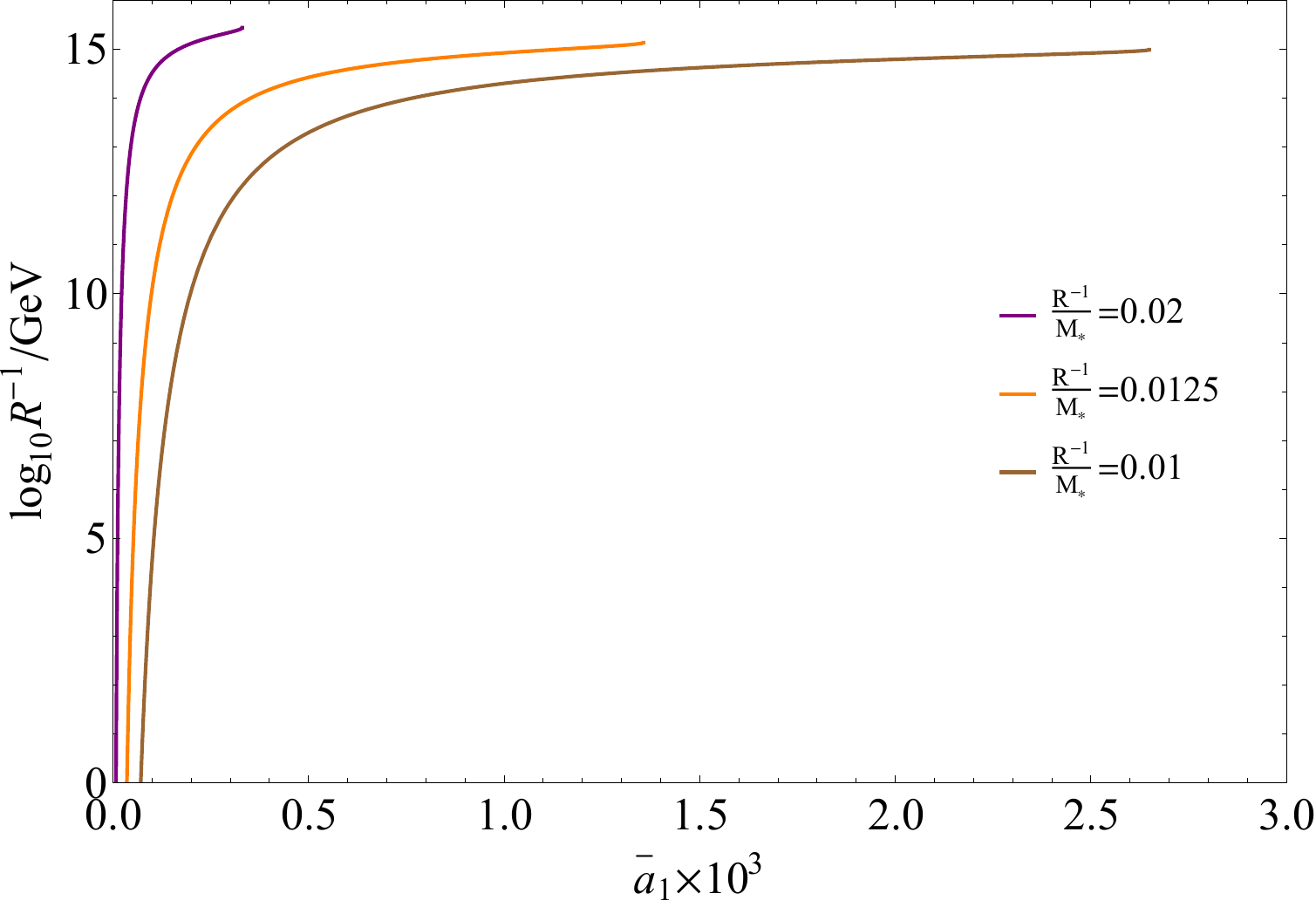}
\end{tabular}
  \caption{The dependence of the inverse size $R^{-1}$ of extra dimension in the parameter $\bar{a}_1$ for various reference points of ratio $R^{-1}/M_*$ with $\lambda_{\widetilde{T}}=7.32\times10^{-13}$.}\label{Inf-f}
\end{figure}

In Table \ref{Inftab}, the effective coupling $\lambda_{\widetilde{T}}$ is determined from the observed value of the scalar perturbation amplitude $A_s$. 
For the number of e-foldings $N\sim60$, the effective coupling $\lambda_{\widetilde{T}}$ is around $\sim7\times10^{-13}$ which is very small. In order to see the smallness of the effective coupling $\lambda_{\widetilde{T}}$, let us first rewrite $\lambda_{\widetilde{T}}$ in as
\begin{eqnarray}
\lambda_{\widetilde{T}}&=&\frac{\bar{a}_1}{576}\frac{R^{-2}}{M^2_{\text{Pl}}}\left[\int^\pi_{-\pi}d\theta\chi^2(\theta)\right]\left[\int^\pi_{-\pi}d\theta\chi(\theta)\right]^{-1},\label{inf-lamT}
\end{eqnarray}
where $\bar{a}_1\equiv a_1\langle T\rangle^3$ and $\chi(\theta)$ corresponds to the solution (\ref{negLacc1}). This relation implies that the effective coupling $\lambda_{\widetilde{T}}$ would be suppressed by the observed Planck scale. By using Eq. (\ref{4DPlscal}), we can determine the relation between the parameter $\bar{a}_1$ and the inverse size of extra dimension for the ratio $R^{-1}/M_*$ kept fixed. Such a relation is shown in Fig. \ref{Inf-f}. We observe that  $R^{-1}$ increases with the growth of $\bar{a}_1$, which suggests the large hierarchy between $M_*$, $R^{-1}$ and the electroweak scale in the region of the sufficiently large/small $\bar{a}_1$. Otherwise, no the large hierarchy appears in the intermediate region of $\bar{a}_1$.  In addition, for $R^{-1}/M_*$ given, $\bar{a}_1$ and $R^{-1}$ cannot exceed a certain value, which is due to the behavior of the function $\chi(\theta)$ and the constraints coming from the identification of the observed Planck scale, and the observed value of the scalar perturbation amplitude. For example, with $R^{-1}/M_*=0.0125$ corresponding to the orange curve in Fig. \ref{Inf-f}, we find $\bar{a}_1\lesssim1.356\times10^{-3}$ and $R^{-1}\lesssim1.41\times10^{15}$ GeV.

At the end of the inflation, the inflaton oscillates around the global minimum or VEV $\langle\chi\rangle$ and corresponds to a mass
\begin{eqnarray}
m_\chi&=&16\sqrt{\frac{3}{5}\lambda_{\widetilde{T}}}M_{\text{Pl}}.
\end{eqnarray}
With the number of e-foldings $N\sim60$, one finds that the mass of the inflaton is about $\sim2.6\times10^{13}$ GeV.

\section{\label{Geoeq}Geodesic equation and Observations}
In this section, we shall derive the geodesic equation which describes the motion of neutral test particles in spacetime $\mathcal{M}_5$. Then, we consider the phenomenological implications of the geodesic equation for the motion of neutral test particles. We are interested in investigating the photon sphere and the bound orbits, which have the important differences between GR and our scenario about the qualitative and quantitative features and hence would allow us to distinguish our scenario from other alternative theories of gravity. In addition, we calculate the perihelion shift and the bending angle of light in the weak field limit and then we compare the values predicted by our spacetime $\mathcal{M}_5$ with the experimental observations to find the corresponding constraints. With respect the gravitational deflection of light in the strong field limit, we study the impact of spacetime $\mathcal{M}_5$ on the strong gravitational lensing.

The worldline $\{x^\mu(\lambda),\theta(\lambda)\}\equiv X^M(\lambda)$ of a neutral test particle moving in spacetime $\mathcal{M}_5$ satisfies the geodesic equation
\begin{eqnarray}
\frac{d^2X^M}{d\lambda^2}+\Gamma^M_{PQ}\frac{dX^P}{d\lambda}\frac{dX^Q}{d\lambda}=0,
\end{eqnarray}
where $\lambda$ is the Affine parameter. In addition, the line element of spacetime $\mathcal{M}_5$ implies the following relation
\begin{eqnarray}
\bar{g}_{\mu\nu}(x)\frac{dx^\mu}{d\lambda}\frac{dx^\nu}{d\lambda}=\frac{1}{\chi(\theta)}\left[k+\frac{\widetilde{T}^4}{\Delta^2}\left(\frac{d\theta}{d\lambda}\right)^2\right],\label{met-vel}
\end{eqnarray}
where $k=0(1)$ for the null(timelike) geodesic and
\begin{eqnarray}
\Delta&\equiv&24M^3_*\int^\pi_{-\pi}d\theta\chi(\theta).
\end{eqnarray}
The geodesic equation is more specifically expanded as
\begin{eqnarray}
\frac{dv^\mu}{d\lambda}+\bar{\Gamma}^\mu_{\nu\rho}v^\nu v^\rho+\frac{g_{_X}\chi'(\theta)}{2\chi(\theta)}\left[\frac{X^\mu}{\chi(\theta)}\left(k+\frac{\widetilde{T}^4}{\Delta^2}(v^\theta)^2\right)-2v^\mu v^\nu X_\nu\right]&&\nonumber\\
+\left[\frac{\chi'(\theta)}{\chi(\theta)}{\delta^\mu}_\nu+\frac{g_{_X}\widetilde{T}^4}{\Delta^2\chi(\theta)}{X^\mu}_\nu\right]v^\nu v^\theta+\frac{2\widetilde{T}^3\partial^\mu\widetilde{T}}{\Delta^2\chi(\theta)}(v^\theta)^2&=&0,\label{x-geodeq}\\
\frac{dv^\theta}{d\lambda}+\frac{\chi'(\theta)}{2\chi(\theta)}\left[\frac{\Delta^2}{\widetilde{T}^4}k+(v^\theta)^2\right]-4\frac{\partial_\mu\widetilde{T}}{\widetilde{T}}v^\mu v^\theta&=&0,\label{theta-geodeq}
\end{eqnarray}  
where $v^\mu\equiv dx^\mu/d\lambda$, $v^\theta\equiv d\theta/d\lambda$, 
\begin{eqnarray}
\bar{\Gamma}^\mu_{\nu\rho}&=&\frac{\bar{g}^{\mu\lambda}(x)}{2}\left[\partial_\nu\bar{g}_{\lambda\rho}(x)+\partial_\rho\bar{g}_{\lambda\nu}(x)-\partial_\lambda\bar{g}_{\nu\rho}(x)\right],
\end{eqnarray}
$X^\mu\equiv\bar{g}^{\mu\nu}(x)X_\nu$, ${X^\mu}_\nu\equiv\bar{g}^{\mu\rho}(x)X_{\rho\nu}$, and $\partial^\mu\equiv\bar{g}^{\mu\nu}(x)\partial_\nu$.

For the spherically-symmetric and static solution of the system described by the effective action (\ref{effSPact}), the ansatz for the effectively four dimensional metric $\bar{g}^{\mu\nu}(x)$, the gauge field $X_\mu$, and the field $\widetilde{T}$ are given as
\begin{eqnarray}
\bar{g}_{\mu\nu}(x)dx^\mu dx^\nu&=&f(r)dt^2-\frac{dr^2}{g(r)}-r^2(d\varphi_1^2+\sin^2\varphi_1 d\varphi_2^2),\nonumber\\
X_\mu dx^\mu&=&\phi(r)dt.\nonumber\\
\widetilde{T}&=&\psi(r).\label{SSSB}
\end{eqnarray} 
It should be noted that the massive objects in astronomy are in fact rotating and thus the spacetime geometry of these objects are described by the Kerr metric \cite{Kerr1963}. In this work, we consider the massive objects to be the Sun and black holes. The Sun rotates very slowly with its rotation rate varying between approximately $13.8/$year to $7.7/$year at the equator and the poles, respectively \cite{Beck2000}. Whereas, most black holes are born to have the rotation rate to be very slowly with the rotation parameter $a\equiv J/M^2$ (with $J$ and $M$ to be the anglular momentum and the mass of black hole) about $10^{-2}$ \cite{Fuller2019}. For black holes which are compact components in low-mass X-ray binaries (LMXBs) which are of the most important astrophysical laboratories, all current methods of spin measurement of these black holes need to employ the accretion disk to infer their spin rate. Due to the difficulty in modeling accretion disk, the spin rate measurement sometimes leads to conflicting results. On the other hand, the origin of the black hole spin in LMXBs can be accounted for by the matter accreted onto the black hole \cite{Fragos2015}. In addition, if the accreation is chaotic then the matter accreated on the black holes may not make increasing the black hole spin rate \cite{Miler2014}. With respect to non-accreting stellar-mass black holes whose growth is through mergers with other black holes, successive additions of random spins of merging black holes tend to spin-down the resultant black hole, unless alignment of their spin is very efficient \cite{Berti2008}. The spin measurements of these black holes which are now available for mergers detected by LIGO have pointed to that most of them are consistent with the low spin \cite{Roulet2018,LIGO2018}.

Because of $a\ll1$ in these situations, we can consider the zero order in the rotation parameter $a$, which corresponds to the no rotation approximation of very slowly rotating massive objects.

With the ansatz (\ref{SSSB}), Eqs. (\ref{x-geodeq}) and (\ref{theta-geodeq}) become 
\begin{eqnarray}
\frac{dv^0}{d\lambda}+\frac{f'(r)}{f(r)}v^0v^1+\frac{g_{_X}\chi'(\theta)}{2\chi(\theta)}\phi(r)\left[\frac{1}{\chi(\theta)}\left(k+\frac{\psi^4(r)}{\Delta^2}(v^\theta)^2\right)-2(v^0)^2\right]&&\nonumber\\
+\left[\frac{\chi'(\theta)}{\chi(\theta)}v^0-\frac{g_{_X}\psi^4(r)}{\Delta^2\chi(\theta)}\frac{\phi'(r)}{f(r)}v^1\right]
v^\theta&=&0,\nonumber\\
\frac{dv^1}{d\lambda}+\frac{g(r)f'(r)}{2}(v^0)^2-\frac{g'(r)}{2g(r)}(v^1)^2-rg(r)\left[(v^2)^2+\sin^2\varphi_1(v^3)^2\right]-\frac{g_{_X}\chi'(\theta)}{\chi(\theta)}\phi(r)v^0v^1&&\nonumber\\
+\left[\frac{\chi'(\theta)}{\chi(\theta)}v^1-\frac{g_{_X}\psi^4(r)}{\Delta^2\chi(\theta)}g(r)\phi'(r)v^0\right]
v^\theta-\frac{2g(r)}{\Delta^2\chi(\theta)}\psi^3(r)\psi'(r)(v^\theta)^2&=&0,\nonumber\\
\frac{dv^2}{d\lambda}+\frac{2}{r}v^1v^2-\frac{\sin2\varphi_1}{2}(v^3)^2-\frac{g_{_X}\chi'(\theta)}{\chi(\theta)}\phi(r)v^0v^2+\frac{\chi'(\theta)}{\chi(\theta)}v^2v^\theta&=&0,\nonumber\\
\frac{dv^3}{d\lambda}+\frac{2}{r}v^1v^3+\cot\varphi_1 v^2v^3-\frac{g_{_X}\chi'(\theta)}{\chi(\theta)}\phi(r)v^0v^3+\frac{\chi'(\theta)}{\chi(\theta)}v^3v^\theta&=&0,\nonumber\\
\frac{dv^\theta}{d\lambda}+\frac{\chi'(\theta)}{2\chi(\theta)}\left[\frac{\Delta^2}{\psi^4(r)}k+(v^\theta)^2\right]-\frac{4\psi'(r)}{\psi(r)}v^rv^\theta&=&0.\label{ged-eqs}
\end{eqnarray} 
From the equation for $dv^2/d\lambda$, one sees that we always restrict the equatorial motion of test particles, i.e. $\varphi_1=\pi/2$, without loss of generality.

We consider the propagation of neutral test particles in the low energy region. This implies $\widetilde{T}\simeq\langle\widetilde{T}\rangle$ because as indicated above the fluctuation field around the vacuum $\langle\widetilde{T}\rangle$ is very massive and thus decouples at the low energy. Also, determining the analytical solution for these geodesic equations with $\chi(\theta)$ which is nontrivial is a difficult task. Thus in the following we consider the trivial solution of $\chi(\theta)$ which is given by Eq. (\ref{trisol}). As a result, the spherically-symmetric and static spacetime background given in Eq. (\ref{SSSB}) reads
\begin{eqnarray}
f(r)=g(r)&=&1-\frac{2M}{r}+\frac{Q^2_X}{r^2},\nonumber\\
\phi(r)&=&-\frac{Q_X}{r},\nonumber\\
\psi(r)&=&\langle\widetilde{T}\rangle,\label{SSSP}
\end{eqnarray}
where $M$ is the ADM mass of the system and $Q_X$ is the electric-like charge corresponding to the gauge field $X_\mu$. Because the charge $Q_X$ would be small, we safely neglect the contribution of the term $Q^2_X/r^2$ compared to the term $2M/r$ in the expression of $f(r)$ and $g(r)$. The Schwarzschild radius thus is approximately given by $2M$. Note that, here (and in whats follows) we have considered the unit system where the observed gravitational constant $G$ is given as $G=1$.

Then, by integrating the last expression in Eq. (\ref{ged-eqs}), one finds
\begin{eqnarray}
v^\theta=\frac{c_\theta}{R},\label{verve}
\end{eqnarray}
where $c_\theta$ is a constant which means that the velocity along the vertical direction is always a constant. Substituting this result into the first and fourth lines of Eq. (\ref{ged-eqs}) and then integrating them, we obtain
\begin{eqnarray}
v^0&=&\frac{1}{f(r)}\left[E+\frac{c_\theta}{\sqrt{4\pi}}\phi(r)\right],\nonumber\\
v^3&=&\frac{L}{r^2},\label{v0v3Eq}
\end{eqnarray}
where $E$ and $L$ are the constants which are at large distance identified as the specific energy and the specific angular momentum of the massive particles (while for the massless particles like the photon $E$ and $L$ can be thought of as the energy and angular momentum, respectively) measured by a stationary observer at large radius. With this result and using Eq. (\ref{met-vel}), we obtain
\begin{eqnarray}
(v^1)^2=\left[E+\frac{c_\theta}{\sqrt{4\pi}}\phi(r)\right]^2-f(r)\left(k+c^2_\theta+\frac{L^2}{r^2}\right).\label{v1eq}
\end{eqnarray}
This equation can be rewritten as
\begin{eqnarray}
(v^1)^2=\left(E-V_-\right)\left(E-V_+\right),
\end{eqnarray}
where the effective potentials $V_{\pm}$ are defined as follows
\begin{eqnarray}
V_{\pm}=-\frac{c_\theta}{\sqrt{4\pi}}\phi(r)\mp\sqrt{f(r)\left(k+c^2_\theta+\frac{L^2}{r^2}\right)}.
\end{eqnarray}

\subsection{Photon Sphere}

Eq. (\ref{v1eq}) is written for the photon ($k=0$) which moves along the null paths in the background given in (\ref{SSSP}) as
\begin{eqnarray}
\left(\frac{dr}{d\lambda}\right)^2=\left(E+\frac{\alpha}{r}\right)^2-\left(1-\frac{2M}{r}\right)\left(c^2_\theta+\frac{L^2}{r^2}\right),\label{pteq}
\end{eqnarray}
where
\begin{eqnarray}
\alpha\equiv-\frac{c_\theta Q_X}{\sqrt{4\pi}},\label{alpdef}
\end{eqnarray}
which is positive or negative depending on the sign of the charge $Q_X$. Here, we observe that the structure of spacetime $\mathcal{M}_5$ affects the orbital dynamics of the photon through the additional terms related to the quantities $\alpha$ and $c^2_\theta$. In order to see their effects on the orbital dynamics of the photon, we investigate separately their role corresponding to two cases which are $c^2_\theta\simeq0$ and $\alpha\simeq0$ (or $Q_X\simeq0$).

The conditions for the photon moving on a circular orbit are that the right hand side of Eq. (\ref{pteq}) and its first-order derivative in $r$ vanish both. For the first case of $c_\theta\simeq0$ which hence we ignore the term relating to $c^2_\theta$ in Eq. (\ref{pteq}) and then we find
\begin{eqnarray}
L_\pm&=&\pm r\frac{\sqrt{1-2M/r}}{r-3M}\alpha,\nonumber\\
E_\pm&=&-\frac{\alpha}{r}\pm L_\pm\frac{\sqrt{f(r)}}{r}\nonumber\\
&=&\frac{M\alpha}{r(r-3M)},\label{EL-PS}
\end{eqnarray}
We note that the energy and the angular momentum of the photon are positive both. First, with respect to the angular momentum of the photon, the positive condition corresponding to the positive branch $L_+$ suggests the constraint on $r$ and $\alpha$ as $r>3M$ and $\alpha>0$ or an alternative $r<3M$ and $\alpha<0$. Corresponding to the negative branch $L_-$, the positive condition leads to $r>3M$ and $\alpha<0$ or $r<3M$ and $\alpha>0$. Whereas, applying the positive condition on the energy of the photon, one finds the same constraint as the case of the positive branch for the angular momentum. This means that only the positive branch $L_+$ of the angular momentum is suitable to the expression of energy.

\begin{figure}[t]
 \centering
\begin{tabular}{cc}
\includegraphics[width=0.6 \textwidth]{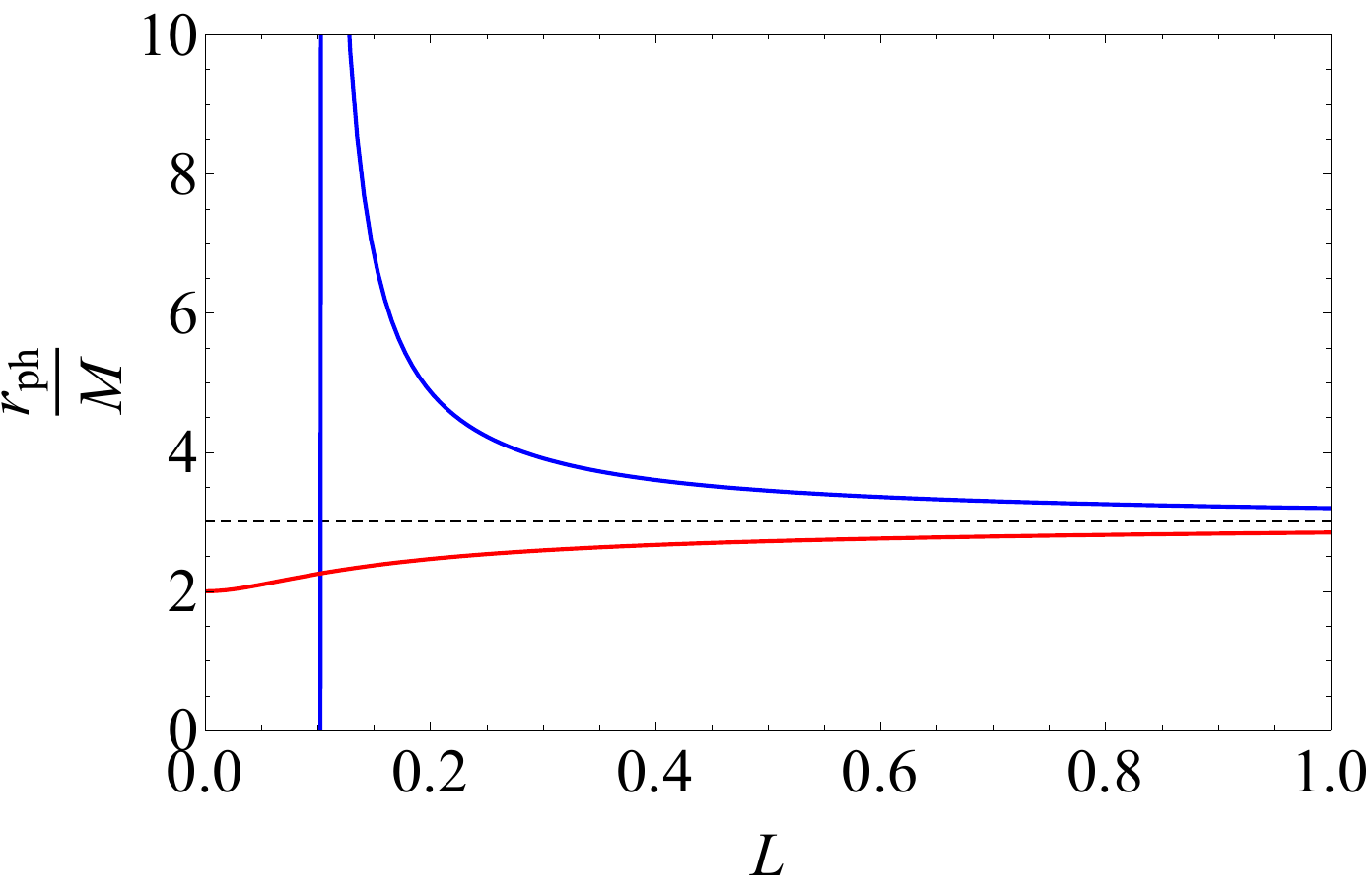}
\end{tabular}
\caption{The scaled radius $r_{ph}/M$ for the photon sphere as a function of the angular momentum $L$ at $|\alpha|=1/10$. The blue and red curves correspond to $r^+_{ph}$ ($\alpha=1/10$) and $r^-_{ph}$ ($\alpha=-1/10$), respectively. The horizontal dashed black line refers to the prediction of GR, $r_{ph}=3M$.}\label{PHfig}
\end{figure}

By solving the first equation (corresponding to $L_+$) in Eq. (\ref{EL-PS}), one determines the radius for the photon sphere as 
\begin{eqnarray}
r^+_{ph}&=&\frac{3L^2/\alpha^2-1+\sqrt{1+3L^2/\alpha^2}}{L^2/\alpha^2-1}M,\nonumber\\
r^-_{ph}&=&\frac{3L^2/\alpha^2-1-\sqrt{1+3L^2/\alpha^2}}{L^2/\alpha^2-1}M.\label{PSrad}
\end{eqnarray}
Here, the positive and negative branches of the solution are always larger and smaller than $3M$ and hence would correspond to $\alpha>0$ and $\alpha<0$, respectively. We show the behavior of the radius $r_{ph}$ for the photon sphere as a function of the angular momentum $L$ in Fig. \ref{PHfig}.

Let us now analyze the features of the radius $r_{ph}$ for the photon sphere derived above.

\begin{itemize}
\item[(1)] In GR and almost alternative theories of gravity \cite{Bahamonde2019,GuoMiao2020,Boonserm2020,Turimov2020,XXZeng2020,GuoLi2020} as well as references therein, the radius $r_{ph}$ for the photon sphere is fixed by the geometry of spacetime. However, in our work, besides the dependence of the geometry of spacetime $r_{ph}$ is also dependent on the angular momentum $L$ of the photon. 

\item[(2)] For the case $|\alpha|\ll L$, the radius $r_{ph}$ for the photon sphere given in Eq. (\ref{PSrad}) is expanded as
\begin{eqnarray}
r_{ph}=3M\left[1+\frac{\alpha}{\sqrt{3}L}+\mathcal{O}(\alpha^2/L^2)\right].
\end{eqnarray}
It means that, in the limit of the large angular momentum of the photon, the radius $r_{ph}$ for the photon sphere reduces to $3M$ which is the corresponding value for the Schwarzschild spacetime geometry and hence our prediction for $r_{ph}$ is the same as that of GR. This can be seen in Fig. \ref{PHfig}.

\item[(3)] Our prediction for $r_{ph}$ deviates significantly from GR about the qualitative and quantitative features in the region of the small angular momentum of photon at which the magnitude of the quantity $\alpha$ can be comparable to $L$ or is much larger than $L$, as seen in Fig. \ref{PHfig}.

\item[(4)] The positive branch $r^+_{ph}$ predicts the radius for the photon sphere larger than the prediction of GR. Also, the radius for the photon sphere corresponding to this branch exists only for the region $L\in(\alpha,+\infty)$ (because the radius for the photon sphere is negative for $L<\alpha$) and blows up as $L\rightarrow\alpha$. Whereas, the radius for the photon sphere predicted by the negative branch $r^-_{ph}$ lies in the region $(2M,3M)$ and hence it yields a smaller radius for the photon sphere compared to the prediction of GR.
\end{itemize} 

For the second case of $\alpha\simeq0$, from the conditions for the photon moving on a circular orbit in analogy to what we have done in the first case, we find two radii as
\begin{eqnarray}
r^{\pm}_{ph}&=&\frac{L^2}{2c^2_\theta M}\left(1\pm\sqrt{1-\frac{12c^2_\theta M^2}{L^2}}\right),
\end{eqnarray}
but only if $L^2\geq12c^2_\theta M^2$. The radii $r^+_{ph}$ and $r^-_{ph}$ correspond to the stable and unstable circular orbits, respectively. Because the fact that the quantity $c_\theta$ is small, the radius $r^+_{ph}$ of the stable circular orbit which is not predicted in GR is very large and approaches infinity when $c_\theta$ goes to zero. This means that the stable circular orbit of the photon manifests significantly for $c_\theta$ sufficiently large. Here, we have several comments in order:

\begin{itemize}
\item[(1)] In the case of $L=12c^2_\theta M^2$, the radii $r^+_{ph}$ and $r^-_{ph}$ coincide and in which case it leads to ISCO for the photon with radius $r_c=6M$ which is equal to the ISCO radius of the massive particle in the Schwarzschild geometry.

\item[(2)] For the photon with angular momentum $L$ below $12c^2_\theta M^2$ corresponding to the fact that the effective potential has no minimum and maximum, there is no turning point to hit and since the photon keeps moving and never returns.

\item[(3)] Because the presence of the stable and unstable circular orbits, there exists the orbits for the photon which is bound but non-circular with the suitable energy.
\end{itemize}

\subsection{Bound orbits of massive particles}

First, we analyze the circular orbits of the massive particle by investigating Eq. (\ref{v1eq}) with $k=1$, given as follows
\begin{eqnarray}
\left(\frac{dr}{d\lambda}\right)^2=\left(E+\frac{\alpha}{r}\right)^2-\left(1-\frac{2M}{r}\right)\left(1+c^2_\theta+\frac{L^2}{r^2}\right)\equiv V(r),\label{TMeq}
\end{eqnarray}
where $c_\theta$ must be realized to be the velocity along the vertical direction of the massive particle and the parameter $\alpha$ is defined the same as in Eq. (\ref{alpdef}). Note that, here the value of $\alpha$ is generally different to the photon case due to the difference in the velocity along the vertical direction (relating to $c_\theta$) and the $Q_X$ charge caused by two different sources. The conditions for the massive particle moving on a circular orbit are $V(r)=0$ and $V'(r)=0$, which lead to the following equation
\begin{eqnarray}
(3M-r)L^2+(1+c^2_\theta)Mr^2+\alpha\sqrt{r(r-2M)\left[L^2+(1+c^2_\theta)r^2\right]}=0,\label{COrads}
\end{eqnarray}
which determines radii for the circular orbits of the massive particle. It is not easily to obtain analytical solutions for Eq. (\ref{COrads}). In fact $\alpha$ is small because the predictions of GR agree with the experimental observations, in particular, in the solar system. Hence one can solve perturbatively Eq. (\ref{COrads}), given as
\begin{eqnarray}
r_{\pm}=r_{0\pm}-\frac{\sqrt{r_{0\pm}(r_{0\pm}-2M)\left[L^2+(1+c^2_\theta)r_{0\pm}^2\right]}}{2(1+c^2_\theta)Mr_{0\pm}-L^2}\alpha+\mathcal{O}(\alpha^2),
\end{eqnarray}
where $r_{0\pm}$ are radii of two circular orbits at the leading order corresponding to $\alpha=0$ (or $Q_X=0$) and are given as
\begin{eqnarray}
r_{0\pm}=\frac{L^2}{2(1+c^2_\theta)M}\left(1\pm\sqrt{1-\frac{12M^2(1+c^2_\theta)}{L^2}}\right),
\end{eqnarray}
but only if $L\geq2\sqrt{3(1+c^2_\theta)}M$. The radii $r_-$ and $r_+$ correspond to the unstable and stable circular orbits, respectively. For $\alpha>0$, the radii $r_-$ and $r_+$ are enhanced and lowered by the correction coming from the fibre fabric of spacetime, respectively. On the contrary, 
the correction with $\alpha<0$ makes the radii $r_-$ and $r_+$ decreasing and increasing, respectively.

The radius $r_c$ for ISCO of the massive particle happens when the radii $r_-$ and $r_+$ coincide. As a result, we find radius $r_c$ for ISCO of the massive particle as
\begin{eqnarray}
r_c=6M-2\sqrt{\frac{2}{1+c^2_\theta}}\alpha+\mathcal{O}(\alpha^2),
\end{eqnarray}
at 
\begin{eqnarray}
L=2\sqrt{3(1+c^2_\theta)}M+\sqrt{\frac{2}{3}}\alpha+\mathcal{O}(\alpha^2).
\end{eqnarray}  
Note that, the value $r_c=6M$ is predicted by GR for the Schwarzschild spacetime geometry, which is independent on $c_\theta$ in the case of $\alpha=0$ (or $Q_X=0$). Obviously, the radius $r_c$ for ISCO of the massive particle reduces (enlarges) compared to the prediction of GR if $\alpha>0$ ($\alpha<0$).

\begin{figure}[t]
 \centering
\begin{tabular}{cc}
\includegraphics[width=0.6 \textwidth]{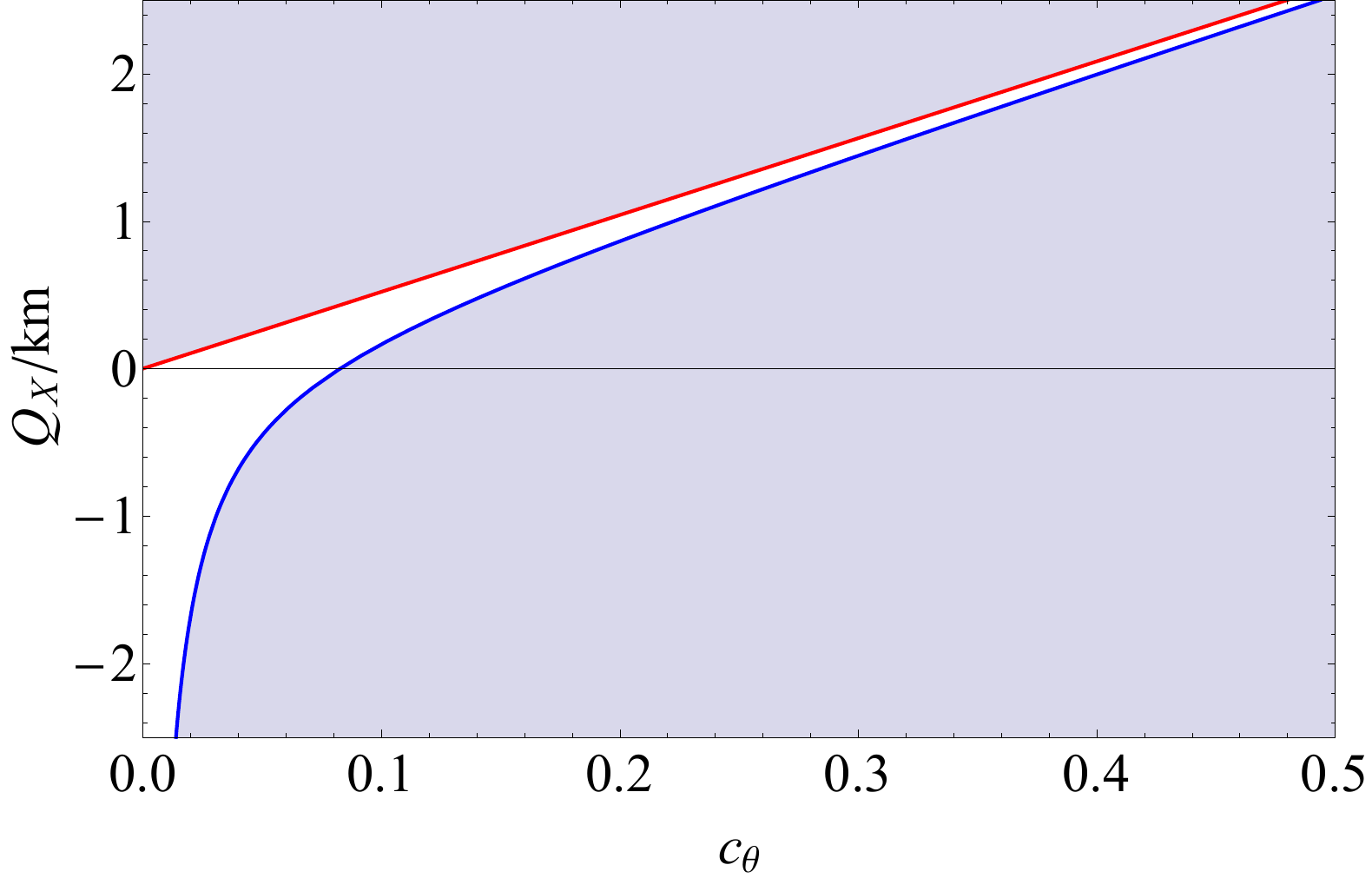}
\end{tabular}
\caption{The allowed parameter region (white region) for the charge $Q_X$ of the Sun and and $c_\theta$ corresponding to Mercury. The red line and blue curve correspond to the upper and lower bounds given in Eq. (\ref{Presh-bs}).}\label{Prehsh}
\end{figure}

We arrive at studying the bound but non-circular orbits for the massive particle. From the second line of Eq. (\ref{v0v3Eq}) and Eq. (\ref{TMeq}), one finds the equation describing the orbital dynamics of a massive particle as
\begin{eqnarray}
\left(\frac{dy}{d\varphi_2}\right)^2&=&\frac{1}{L^2}\left[\left(E+\frac{\alpha}{r_+}\right)^2-\left(1-\frac{2M}{r_+}\right)\left(1+c^2_\theta+\frac{L^2}{r^2_+}\right)\right]+\frac{2}{L^2}\left[\frac{L^2}{r^2_+}(3M-r_+)+M(1+c^2_\theta)\right.\nonumber\\
&&\left.+\alpha\left(E+\frac{\alpha}{r_+}\right)\right]y-\left(1-\frac{6M}{r_+}-\frac{\alpha^2}{L^2}\right)y^2+2My^3,
\end{eqnarray}
where $y\equiv\frac{1}{r}-\frac{1}{r_{+}}$ which describes the deviation from the stable circular orbit. For the orbit which is nearly circular and hence $y$ is small, we can ignore the term in $y^3$ and the solution has the following form
\begin{eqnarray}
y=y_0+A\cos(k\varphi_2+\varphi_{20}),
\end{eqnarray}
where the constants $y_0$, $A$, and $k$ are given as
\begin{eqnarray}
y_0&=&\frac{L^2(3M-r_+)+r^2_+\left[M(1+c^2_\theta)+\alpha(E+\alpha/r_+)\right]}{r_+\left[L^2(r_+-6M)-\alpha^2r_+\right]},\nonumber\\
k&=&\left(1-\frac{6M}{r_+}-\frac{\alpha^2}{L^2}\right)^{1/2},\nonumber\\
A&=&\frac{1}{k}\left\{\frac{1}{L^2}\left[\left(E+\frac{\alpha}{r_+}\right)^2-\left(1-\frac{2M}{r_+}\right)\left(1+c^2_\theta+\frac{L^2}{r^2_+}\right)\right]+k^2y^2_0\right\}^{1/2},\label{Preshpar}
\end{eqnarray}
and $\varphi_{20}$ is the initial phase which can be set to be zero for simplicity. We have two remarks as follows:
\begin{itemize}
\item[(1)] In GR and almost alternative theories of gravity, $y_0$ vanishes because $r_+$ satisfies the equation obtained from the vanishing of the first-order derivative of the effective potential. Thus, the bound orbits which is nearly circular will oscillate around the radius of the stable circular orbit. However, in our case $y_0$ is generally nonzero unless $\alpha=0$ (or $Q_X=0$) and dependent on the specific energy and angular momentum of the particle. As a result, the orbit oscillates not about $y=0$ but about $y=y_0$ which is not fixed by the geometry of spacetime.

\item[(2)] The perihelion shift per orbit is
\begin{eqnarray}
\Delta\varphi_2=\frac{2\pi}{k}-2\pi\simeq\frac{6\pi M}{\hat{r}_{+}}\left(1+\frac{\sqrt{\hat{r}_{+}(\hat{r}_{+}-2M)(L^2+\hat{r}_{+}^2})}{\hat{r}_{+}(2M\hat{r}_{+}-L^2)}\alpha+\frac{M\hat{r}_{+}}{L^2\sqrt{1-12M^2/L^2}}c^2_\theta\right).\label{peshift}
\end{eqnarray}
where
\begin{eqnarray}
\hat{r}_{+}=\frac{L^2}{2M}\left(1+\sqrt{1-\frac{12M^2}{L^2}}\right),
\end{eqnarray}
is the prediction of GR for the radius of the stable circular orbit. The first term in the final expression of Eq. (\ref{peshift}) is the GR contribution, whereas the second and third terms are the corrections coming from spacetime $\mathcal{M}_5$ up to the second order in $c_\theta$. The measured value for the perihelion shift of Mercury is $42.98\pm0.04''$ per a century \cite{CMWill2006}. Using $\hat{r}_{+}\approx5.550\times10^7$ km, $M=M_\odot\approx1.474$ km where $M_\odot$ is the mass of the Sun, and each orbit of Mercury around the Sun takes approximately $87.97$ days, one finds the value as predicted by GR for the perihelion shift of Mercury as $\approx42.75''$ per a century. Then, from the observed and GR values and $L^2\simeq M_\odot\hat{r}_+$, we obtain the constraint on the charge $Q_X$ of the Sun and and $c_\theta$ corresponding to Mercury as
\begin{eqnarray}
\sqrt{4\pi}\left(1.47c_\theta-\frac{10^{-2}}{c_\theta}\right)\lesssim\frac{Q_X}{\text{km}}\lesssim5.2c_\theta.\label{Presh-bs}
\end{eqnarray}
These upper and lower bounds are depicted in Fig. \ref{Prehsh} where the white region refers to the allowed parameter region.
\end{itemize}

\subsection{Gravitational deflection of light}
We determine the deflection angle of light when it passes through a central massive object in our spacetime $M_5$, given as
\begin{eqnarray}
\Delta\varphi_2=2\int^\infty_{r_0}dr\frac{L}{r^2}\left[\left(E+\frac{\alpha}{r}\right)^2-\left(1-\frac{2M}{r}\right)\left(c^2_\theta+\frac{L^2}{r^2}\right)\right]^{-1/2}-\pi,
\end{eqnarray}
where $r_0$ is the point of closest approach to the central massive object which is determined by $dr/d\lambda=0$ corresponding to
\begin{eqnarray}
\left(E+\frac{\alpha}{r_0}\right)^2-\left(1-\frac{2M}{r_0}\right)\left(c^2_\theta+\frac{L^2}{r^2_0}\right)=0.
\end{eqnarray}
In the approximation of the weak deflection, i.e. $\frac{M}{r}\ll1$, and the small corrections coming from spacetime $M_5$, i.e. $\frac{\alpha}{L}\ll1$ or $c_\theta\ll1$, we obtain the deflection angle of light to the subleading order approximation of $c_\theta$ as 
\begin{eqnarray}
\Delta\varphi_2&\simeq&2\left[\left(1+\frac{15M^2}{4b^2}+\frac{2M\alpha}{bL}+\frac{48M^3\alpha}{b^3L}\right)\arcsin(by)-\left(2+\frac{64M^2}{3b^2}+\frac{15My}{4}+\frac{32M^2y^2}{3}
\right.\right.\nonumber\\
&&\left.\left.+\frac{15M\alpha}{bL}+\frac{48M^2\alpha y}{bL}\right)M\sqrt{\frac{1}{b^2}-y^2}\right]\Big|^{y=1/b}_{y=0}-\pi\nonumber\\
&\simeq&\frac{4M}{b}+\frac{15\pi M^2}{4b^2}+\frac{128M^3}{3b^3}+2\left(\pi+\frac{15M}{b}+\frac{24\pi M^2}{b^3}\right)\frac{M\alpha}{bL},
\end{eqnarray}
where $y$ is related to $r$ as $y^2\equiv\frac{1}{r^2}\left(1-\frac{2M}{r}\right)-\frac{\alpha}{bLr}$ and $b\equiv\frac{L}{E}$ is the impact parameter. The first three terms in the second line are the prediction of GR where the first term $4M/b$ is the leading contribution or the standard GR light deflection angle which is usually considered in the literature \cite{GRbook-Schutz}. Whereas, the terms concerning $\alpha/L$ are the corrections coming from spacetime $\mathcal{M}_5$. We note here that, in obtaining the above deflection angle of light, we have expanded up to $\frac{M^3}{r^3}$ to compare the subleading contributions of GR with the the $\alpha/L$ contributions. This should allow to impose the strong constraint on the corrections coming from our model, as seen later.

The ratio between the observed value of light deflection angle and the corresponding GR prediction has been found as \cite{Shapiro2004,Will2014}
\begin{eqnarray}
\frac{\left(\Delta\varphi_2\right)_{\text{exp.}}}{\left(\Delta\varphi_2\right)_{\text{GR}}}\approx1.0001\pm0.0001.
\end{eqnarray}
In solar system, the gravitational deflection of radio signals has been measured using the Very Long Baseline Interferometry (VLBI) \cite{Robertson1991}. With the numerical values, $M=M_\odot\approx1.474$ km and $b\approx6.96\times10^5$ km, we find the constraint for $\alpha/L$ as
\begin{eqnarray}
0\lesssim\frac{\alpha}{L}\lesssim1.27\times10^{-4}.\label{GDLbound}
\end{eqnarray}

\begin{figure}[t]
 \centering
\begin{tabular}{cc}
\includegraphics[width=0.45 \textwidth]{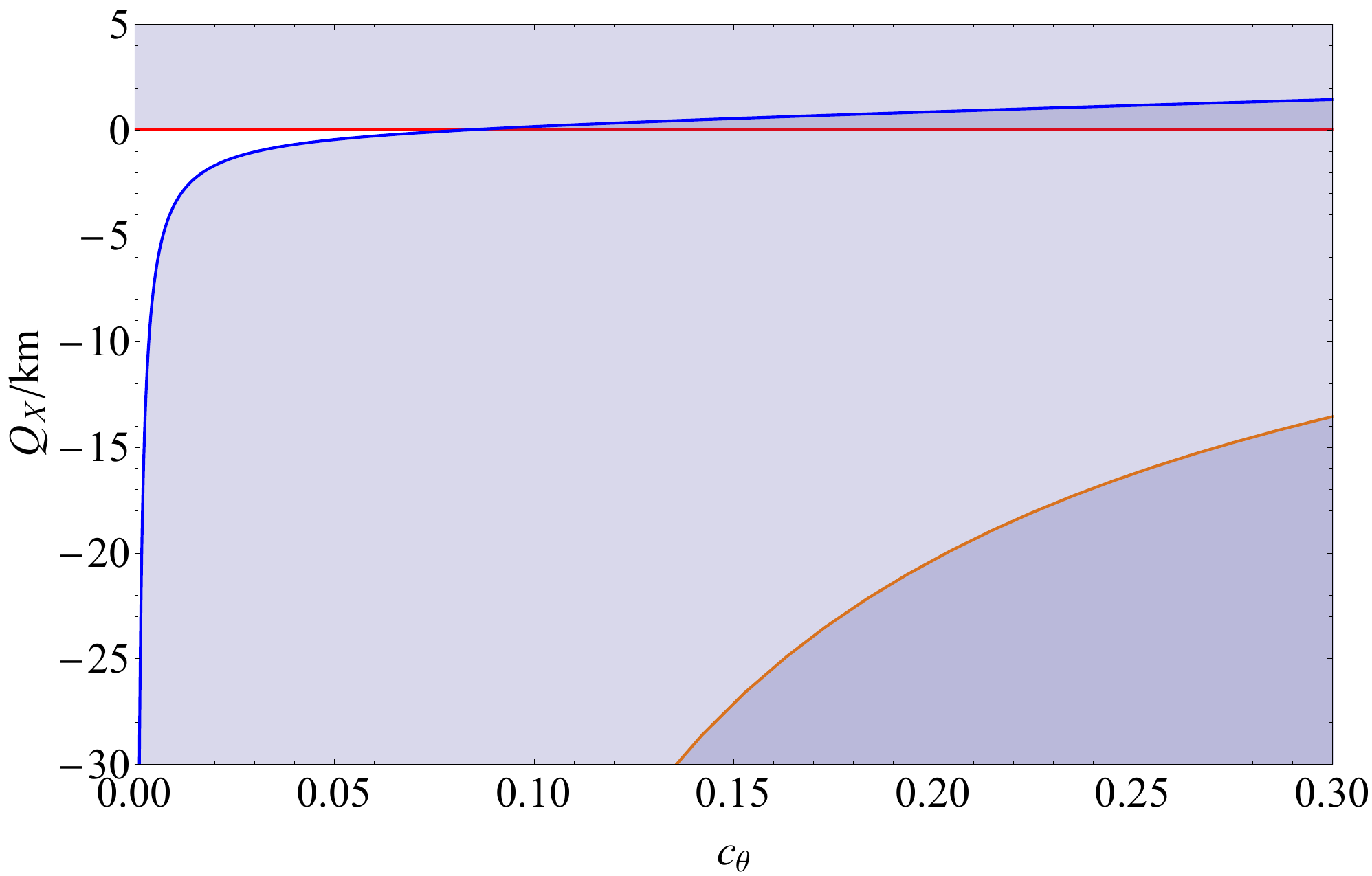}
\hspace*{0.05\textwidth}
\includegraphics[width=0.45 \textwidth]{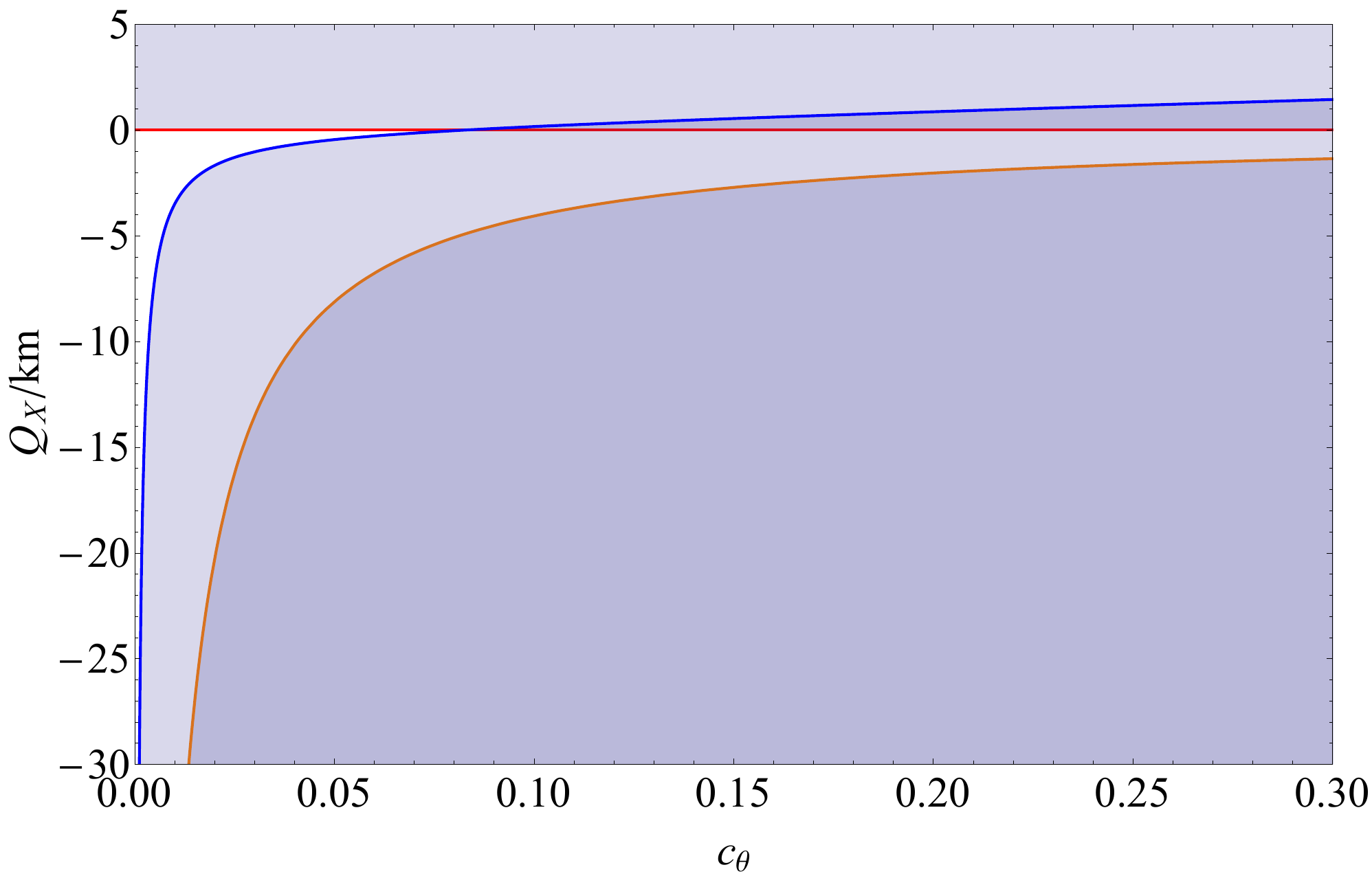}\\
\includegraphics[width=0.45 \textwidth]{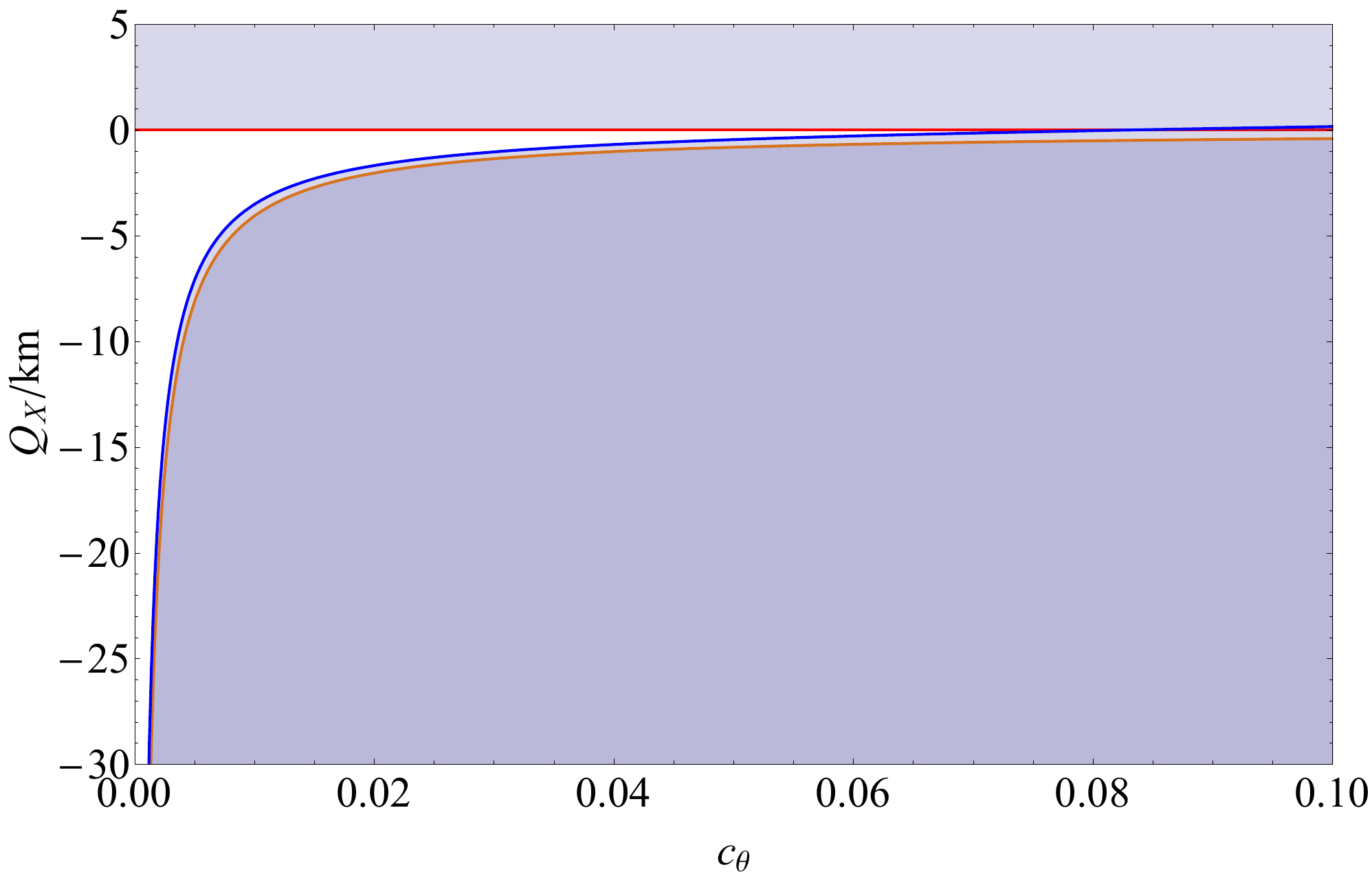}
\hspace*{0.05\textwidth}
\includegraphics[width=0.45 \textwidth]{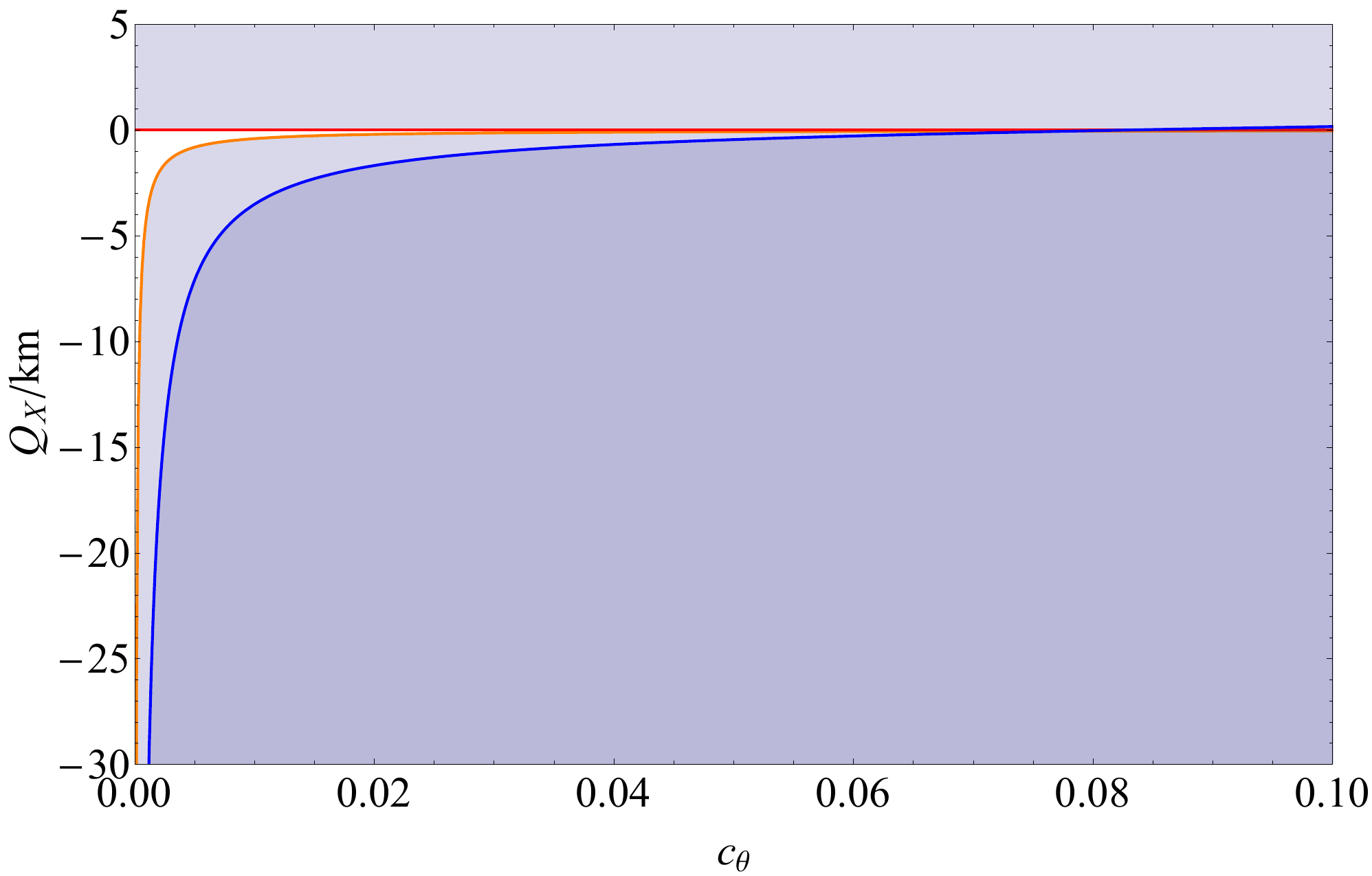}
\end{tabular}
 \caption{The allowed parameter region (white region) for the charge $Q_X$ of the Sun and and the velocity $c_\theta$ of the particles along the direction of the internal space. The red horizontal line and orange curve correspond to the constraint (\ref{GDLbound}), whereas, the blue curve corresponds to the constraint (\ref{Presh-bs}). Top-left panel: $L=\sqrt{M_\odot\hat{r}_+}\approx9032.44$ km. Top-right panel: $L=0.1\sqrt{M_\odot\hat{r}_+}$. Bottom-left panel: $L=0.01\sqrt{M_\odot\hat{r}_+}$. Botom-right panel: $L=0.001\sqrt{M_\odot\hat{r}_+}$.}\label{ComExpbs}
\end{figure}

Assume that the velocity of the massive particles and the photon along the vertical direction or the direction of the internal space is the same order, we combine the experimental bounds given by Eqs. (\ref{Presh-bs}) and (\ref{GDLbound}), which is shown in Fig. \ref{ComExpbs} for various values of the angular momentum $L$ of the photon.  We observe that the constraint of the light bending angle becomes more strongly with the decreasing of the angular momentum of the photon. When the angular momentum of the photon is low enough compared to the specific angular momentum of the massive particles $L=\sqrt{M_\odot\hat{r}_+}$, the constraint of the light bending angle is more stringent than that of the perihelion shift. On the contrary, the allowed parameter region in the $c_\theta-Q_X$ plane is determined by the constraint coming from the perihelion shift.

\subsection{Strong gravitational lensing}
\begin{figure}[t]
 \centering
\begin{tabular}{cc}
\includegraphics[width=0.6 \textwidth]{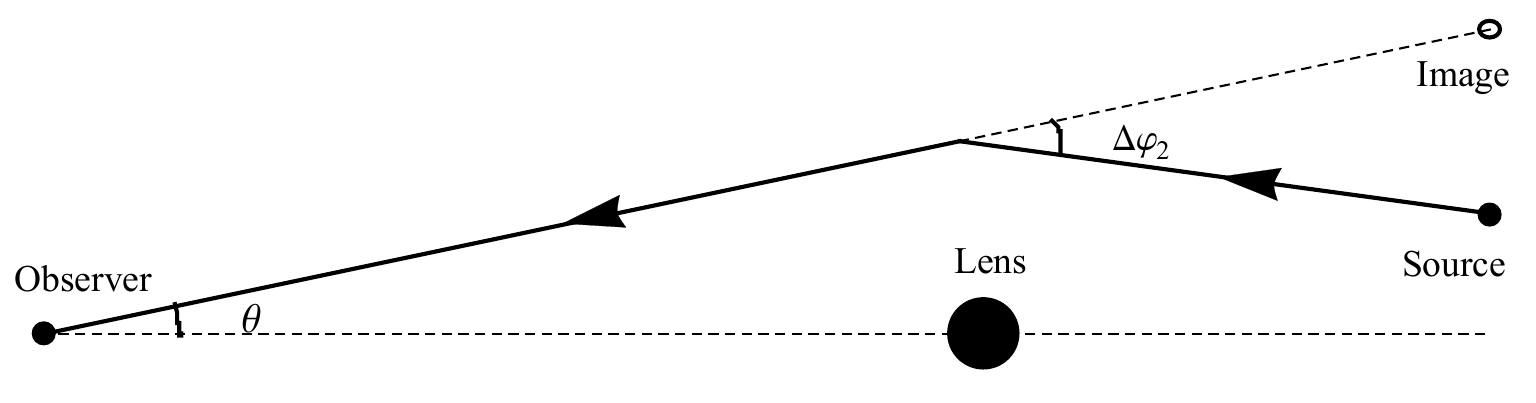}
\end{tabular}
\caption{The diagram of the gravitational lensing.}\label{Gralen-Diag}
\end{figure}

In this subsection, we study the impact of spacetime $\mathcal{M}_5$ on the observables of the strong gravitational lensing whose diagram is given in Fig. \ref{Gralen-Diag}.
The gravitational lens is considered in this work to be the supermassive black hole (SMBH) of Sagittarius A$^*$ at the center of our galaxy (Milky Way) and the galaxies which are Andromeda and Messier 87 (or Virgo A). The observable data about the mass of Sagittarius A$^*$ SMBH and Andromeda and Messier 87 galaxies and the distance (denoted $D_{OL}$) between them and Earth (the observer) are given in Table \ref{SMBH-obda}.
\begin{table}[!htp]
\centering
\begin{tabular}{ccc}
  \hline
  \hline
  \ \ \ \ \text{Gravitational lens}\ \  \ \ & Mass $(M_\odot)$ & $D_{OL}$ (Mpc)  \\
  \hline
  \hline
  Sagittarius A$^*$ \cite{Ghez2008}& $\ \ \ \ \ \ 4.1\times10^6 \ \ \ \ \ \ $ & $\ \ \ \ \ \ 8\times10^{-3} \ \ \ \ \ \ $ \\
  \hline
  Andromeda \cite{Kafle2018}& $\ \ \ \ \ \ 8.0\times10^{11} \ \ \ \ \ \ $ & $\ \ \ \ \ \ 0.78 \ \ \ \ \ \ $ \\
  \hline
  Messier 87 \cite{Murphy2011}& $\ \ \ \ \ \ 5.7\times10^{12} \ \ \ \ \ \ $ & $\ \ \ \ \ \ 16.4 \ \ \ \ \ \ $ \\
  \hline
  \hline
\end{tabular}
\caption{The mass of the Sagittarius A$^*$ SMBH and the Andromeda and Messier galaxies and the distance from them to the Earth.} \label{SMBH-obda}
\end{table}

The deflection angle of light as a function of the closest approach distance $r_0$ is rewritten as follows
\begin{eqnarray}
\Delta\varphi_2(r_0)=I(r_0)-\pi,
\end{eqnarray}
where $I(r_0)$ is given as \cite{Virbhadra1998}
\begin{eqnarray}
I(r_0)=\int^\infty_{r_0}\frac{2\sqrt{B(r)}dr}{\sqrt{C(r)}\sqrt{\frac{C(r)}{C_0}\frac{A_0}{A(r)}-1}},
\end{eqnarray}
with
\begin{eqnarray}
A(r)&=&\left(1-\frac{2M}{r}\right)\left(1+\frac{c^2_\theta r^2}{L^2}\right)\left(1+\frac{\alpha b}{Lr}\right)^{-2},\nonumber\\
B(r)&=&\frac{1}{A(r)}\left(1+\frac{\alpha b}{Lr}\right)^{-2},\nonumber\\
C(r)&=&r^2,
\end{eqnarray}
and $A_0\equiv A(r_0)$, $C_0\equiv C(r_0)$. The deflection angle of light becomes large when the closest approach distance $r_0$ decreases. When $r_0$ reaches the radius of the photon sphere, the deflection angle of light is divergent (which is logarithmic in nature) or in other words the light is captured by the supermassive body. In the strong field limit, the closest approach distance $r_0$ of light is very close to the radius of the photon sphere. This means that in this situation one can expand the deflection angle of light around the radius of the photon sphere as \cite{Bozza2002}
\begin{eqnarray}
\Delta\varphi_2(\theta)=-\bar{a}\log\left(\frac{\theta D_{OL}}{b_m}-1\right)+\bar{b}+\mathcal{O}(r_0-r_{ph}),
\end{eqnarray} 
where $\theta$ is the incident angle to the observer, $\bar{a}$ and $\bar{b}$ are the coefficients of the strong gravitational lensing, and $b_m$ is the minimal impact parameter corresponding to the photon sphere. In order to calculate the coefficients $\bar{a}$ and $\bar{b}$, we follow the method developed by Bozza \cite{Bozza2002}. They are given as
\begin{eqnarray}
\bar{a}&=&\frac{R(0,r_{ph})}{2\sqrt{\beta_{ph}}},\nonumber\\
\bar{b}&=&-\pi+b_R+\bar{a}\log\frac{2\beta_{ph}}{A_{ph}},\label{abbar}
\end{eqnarray}
where
\begin{eqnarray}
\beta_{ph}&=&\frac{C_{ph}(1-A_{ph})^2\left[C''(r_{ph})A_{ph}-C_{ph}A''(r_{ph})\right]}{2A^2_{ph}C'(r_{ph})^2},\nonumber\\
b_R&=&\int^1_0\left[R(z,r_{ph})f(z,r_{ph})-R(0,r_{ph})f_0(z,r_{ph})\right]dz,\label{beph-bR}
\end{eqnarray}
with $C_{ph}\equiv C(r_{ph})$ and $A_{ph}\equiv A(r_{ph})$. In Eqs. (\ref{abbar}) and (\ref{beph-bR}), the functions $R(z,r_0)$, $f(z,r_0)$ are defined as follows
\begin{eqnarray}
R(z,r_0)&=&\frac{2\sqrt{BA}}{CA'}(1-A_0)\sqrt{C_0},\nonumber\\
f(z,r_0)&=&\frac{1}{\sqrt{A_0-[(1-A_0)z+A_0]\frac{C_0}{C}}},
\end{eqnarray}
with $z\equiv[A(r)-A_0]/[1-A_0]$ and the function $f_0(z,r_0)$ is the expansion of the argument of the square root in $f(z,r_0)$ to the second order in $z$ and given as
\begin{eqnarray}
f_0(z,r_0)&=&\frac{1}{\sqrt{\alpha z+\beta z^2}},
\end{eqnarray}
where
\begin{eqnarray}
\alpha&=&\frac{1-A_0}{C_0A'(r_0)}\left[C'(r_0)A_0-C_0A'(r_0)\right],\nonumber\\
\beta&=&\frac{(1-A_0)^2}{2C^2_0A'(r_0)^3}\left[2C_0C'(r_0)A'(r_0)^2+(C_0C''(r_0)-2C'(r_0)^2)A_0A'(r_0)\right.\nonumber\\
&&\left.-C_0C'(r_0)A_0A''(r_0)\right].
\end{eqnarray}

Now we study the impact of spacetime $\mathcal{M}_5$ on the observational quantities of the strong gravitational lensing. The angular separation between the outermost and innermost images and the ratio of the magnification of the outermost image to the others can be approximated in the situation that the lens, source, and observer are highly aligned as \cite{Bozza2002}
\begin{eqnarray}
s&=&\theta_\infty e^{\frac{\bar{b}-2\pi}{\bar{a}}},\nonumber\\
\mathcal{R}&=&2.5\log_{10}\left(e^{\frac{2\pi}{\bar{a}}}\right),
\end{eqnarray}
where $\theta_\infty$ represents the asymptotic angular position where all images except the outermost one are packed together and can expressed as $\theta_\infty=b_m/D_{OL}$. Note that, the expression in the logarithm function, i.e. $e^{\frac{2\pi}{\bar{a}}}$, is the ratio of the flux of the outermost image to the others.

In Tables \ref{SagA-case}, \ref{Andro-case}, and \ref{VirgoA-case}, we show the numerical estimation of the observables of the strong gravitational lensing to the subleading order approximation of $c_\theta$ for the realistic cases which are given in Table \ref{SMBH-obda}. Here, the predictions by GR and spacetime $\mathcal{M}_5$ correspond to $\alpha/L=0$ and $\alpha/L\neq0$, respectively. From these tables, we observe that the $\bar{a}$ and $\bar{b}$ coefficients and the relative magnification $\mathcal{R}$ are well-approximately independent on where the light comes from but only depend on the parameter $\alpha/L$. In addition, we find that the angular position $\theta_\infty$ of the innermost image and the angular separation $s$ between the outermost and innermost images grow with the increasing of the parameter $\alpha/L$. This means that the size of Einstein ring becomes bigger with the growth of the parameter $\alpha/L$. Whereas, the relative magnification $\mathcal{R}$ decreases with the increasing of the parameter $\alpha/L$. With the variation of the observables of the strong gravitational lensing, namely $\theta_\infty$, $s$, and $\mathcal{R}$, in the parameter $\alpha/L$, by measuring these quantities and comparing to the predictions of GR, we can determine the corrections of spacetime $M_5$ on the observables of the strong gravitational lensing. This could allow us to detect the existence of spacetime $M_5$ at the more fundamental level.

\begin{table}[!htp]
\centering
\begin{tabular}{cccccc}
  \hline
  \hline
  $\ \ \ \  \frac{\alpha}{L}\times10^4 \ \  \ \ $ & $\ \ \ \  \bar{a} \ \  \ \ $ & $\ \ \ \  \bar{b} \ \  \ \ $ & $\ \ \theta_\infty$ ($\mu$as) $\ \ $ & $\ \ s$ ($\mu$as) $\ \ $ & $\ \ \ \  \mathcal{R} \ \  \ \ $\\
  \hline  
  $0$ & $1$ & $\ \ -0.40023\ \ $ & $26.239$ & $0.032838$ & $6.82188$ \\
  \hline  
  $0.3$ & $ \ \ 1.00002 \ \ $ & $\ \ -0.400226\ \ $ & $26.2404$ & $0.0328437$ & $6.82176$ \\
  \hline  
  $0.6$ & $1.00003$ & $\ \ -0.400223 \ \ $ & $26.2417$ & $0.0328493$ & $6.82165$ \\
  \hline  
  $0.9$ & $1.00005$ & $\ \ -0.400219\ \ $ & $26.2431$ & $0.0328549$ & $6.82153$ \\
  \hline  
  $1.2$ & $1.00007$ & $\ \ -0.400216\ \ $ & $26.2445$ & $0.0328605$ & $6.82141$ \\
  \hline
  \hline
\end{tabular}
\caption{The numerical estimation of the observables of the strong gravitational lensing for the case of Sagittarius A$^*$ SMBH.} \label{SagA-case}
\end{table}
\begin{table}[!htp]
\centering
\begin{tabular}{cccccc}
  \hline
  \hline
  $\ \ \ \  \frac{\alpha}{L}\times10^4 \ \  \ \ $ & $\ \ \ \  \bar{a} \ \  \ \ $ & $\ \ \ \  \bar{b} \ \  \ \ $ & $\ \ \theta_\infty$ (mas) $\ \ $ & $\ \ s$ (mas) $\ \ $ & $\ \ \ \  \mathcal{R} \ \  \ \ $\\
  \hline  
  \hline
  $0$ & $1$ & $\ \ -0.40023\ \ $ & $52.5108$ & $0.0657171$ & $6.82188$ \\
  \hline  
  $0.3$ & $ \ \ 1.00002 \ \ $ & $\ \ -0.400226\ \ $ & $52.5136$ & $0.0657284$ & $6.82176$ \\
  \hline  
  $0.6$ & $1.00003$ & $\ \ -0.400223 \ \ $ & $52.5163$ & $0.0657396$ & $6.82165$ \\
  \hline  
  $0.9$ & $1.00005$ & $\ \ -0.400219\ \ $ & $52.519$ & $0.0657509$ & $6.82153$ \\
  \hline  
  $1.2$ & $1.00007$ & $\ \ -0.400216\ \ $ & $52.5218$ & $0.0657622$ & $6.82141$ \\
  \hline
  \hline
\end{tabular}
\caption{The numerical estimation of the observables of the strong gravitational lensing for the case of Andromeda galaxy.} \label{Andro-case}
\end{table}
\begin{table}[!htp]
\centering
\begin{tabular}{cccccc}
  \hline
  \hline
  $\ \ \ \  \frac{\alpha}{L}\times10^4 \ \  \ \ $ & $\ \ \ \  \bar{a} \ \  \ \ $ & $\ \ \ \  \bar{b} \ \  \ \ $ & $\ \ \theta_\infty$ (mas) $\ \ $ & $\ \ s$ (mas) $\ \ $ & $\ \ \ \  \mathcal{R} \ \  \ \ $\\
  \hline 
  \hline 
  $0$ & $1$ & $\ \ -0.40023\ \ $ & $17.7945$ & $0.0222697$ & $6.82188$ \\
  \hline  
  $0.3$ & $ \ \ 1.00002 \ \ $ & $\ \ -0.400226\ \ $ & $17.7954$ & $0.0222735$ & $6.82176$ \\
  \hline  
  $0.6$ & $1.00003$ & $\ \ -0.400223 \ \ $ & $17.7963$ & $0.0222773$ & $6.82165$ \\
  \hline  
  $0.9$ & $1.00005$ & $\ \ -0.400219\ \ $ & $17.7972$ & $0.0222811$ & $6.82153$ \\
  \hline  
  $1.2$ & $1.00007$ & $\ \ -0.400216\ \ $ & $17.7982$ & $0.022285$ & $6.82141$ \\
  \hline
  \hline
\end{tabular}
\caption{The numerical estimation of the observables of the strong gravitational lensing for the case of Messier 87 galaxy.} \label{VirgoA-case}
\end{table}

\subsection{Retrolensing}
In this subsection, we study the influence of spacetime $\mathcal{M}_5$ on retrolensing of light rays emitted by the Sun, which is another situation of strong gravitational lensing. The gravitational lens in the ordinary situation of strong gravitational lensing is situated between the observer and the light source. However, if one consider the light of the Sun which is reflected by a black hole with photon sphere and then reaches the observer which is placed between the gravitational lens and the light source, it can give rise the images with the deflection angle closer to $(2n+1)\pi$ with $n$ to be nonnegative integers. In this way, the black hole with photon sphere can reflect the light ray like a mirror and is usually so-called retrolens.

\begin{figure}[t]
 \centering
\begin{tabular}{cc}
\includegraphics[width=0.5 \textwidth]{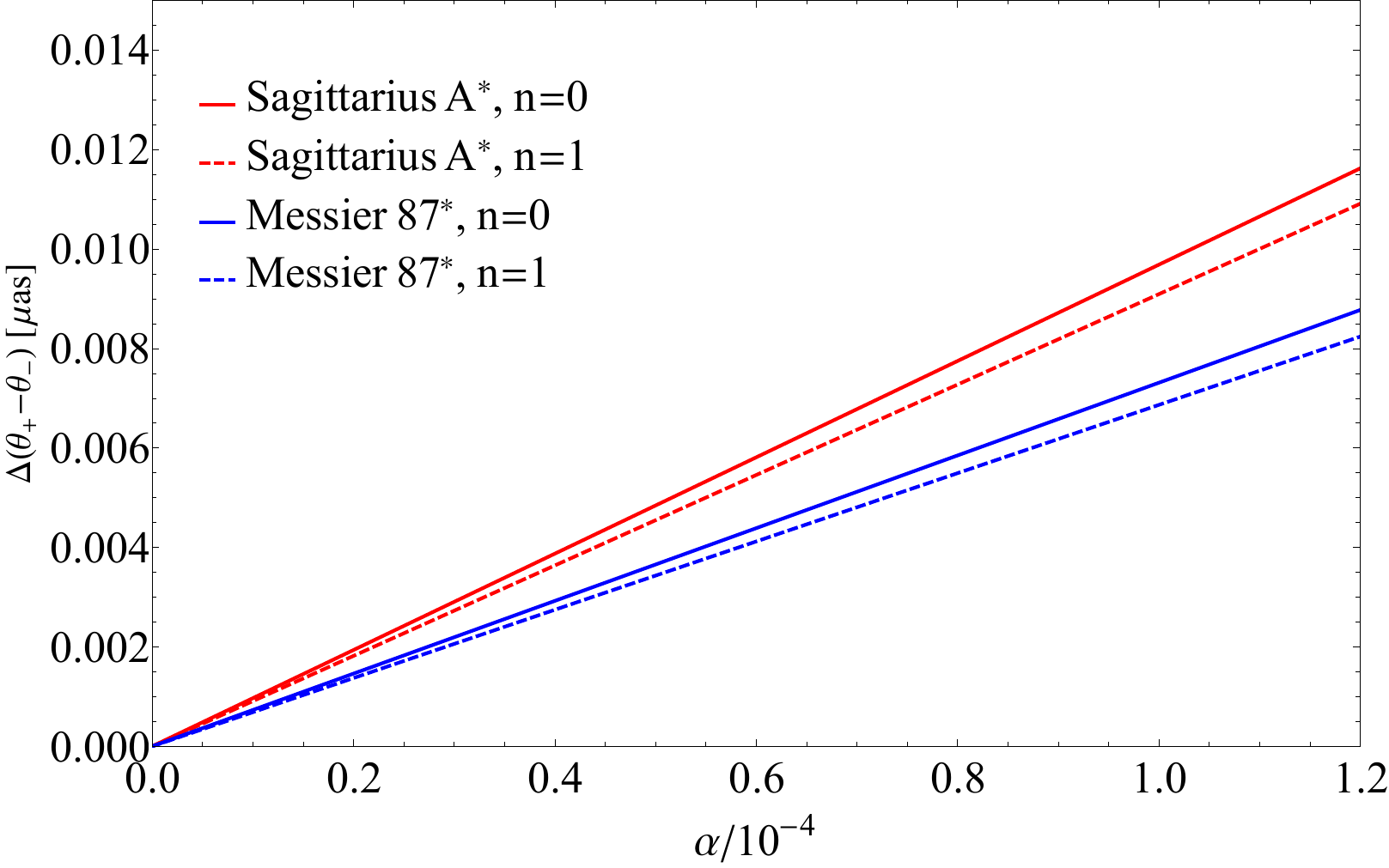}
\end{tabular}
 \caption{The relative difference $\Delta(\theta_+-\theta_-)$ as a function of $\alpha$ for SMBHs of Sagittarius A$^*$ and Messier 87$^*$ as retrolens and the winding number $n=0,1$.}\label{Retr-Delthe}
\end{figure}

In order investigate retrolensing, we introduce the Ohanian lens equation given as \cite{Ohanian1987,Bozza2004}
\begin{eqnarray}
\beta=\pi-\bar{\alpha}(\theta)+\theta+\bar{\theta},
\end{eqnarray}
where the angle $\beta$ is defined as $\beta=\angle OLS$ (with $O$, $L$, $S$ referring to the observer, retrolens, and the Sun, respectively),
$\bar{\alpha}(\theta)=\Delta\varphi_2(\theta)-2\pi n$ is the effective deflection angle with $n$ denoting the winding number of light ray around retrolens before reaching the observer, and $\bar{\theta}$ denotes the angle at $S$ between the light ray emitted from the Sun and the line $LS$. Ignoring the small terms $\theta$ and $\bar{\theta}$, the positive solution of this equation is approximately obtained as
\begin{eqnarray}
\theta_+(\beta)=\frac{b_m}{D_{OL}}\left[1+e^{\frac{\bar{b}-(2n+1)\pi+\beta}{\bar{a}}}\right].
\end{eqnarray}
In addition, retrolens would produce the images in pairs due to the spherical symmetry, called the retrolensing double image, which corresponds to the occurrence of primary and secondary images on the opposite sides of retrolens. On the other hand, there is a negative solution $\theta_-(\beta)=-\theta_+(-\beta)$ of the lens equation. The separation of retrolensing double image is given by
\begin{eqnarray}
\theta_+-\theta_-\simeq2\theta_+=\frac{2b_m}{D_{OL}}\left[1+e^{\frac{\bar{b}-(2n+1)\pi+\beta}{\bar{a}}}\right],
\end{eqnarray}
where the approximation appears when the Sun, observer, retrolens are aligned to be near perfectly, i.e. $\beta\approx0$. We define a quantity $\Delta(\theta_+-\theta_-)$ which is the relative difference between the double image separation $\theta_+-\theta_-$ predicted by our scenario and that predicted by GR. In Fig. \ref{Retr-Delthe}, we show the relative difference $\Delta(\theta_+-\theta_-)$ with Sagittarius A$^*$ and Messier 87$^*$ (which is SMBH at the center of Messier 87 galaxy) as retrolens. Here, the retrolens masses and the distances from Earth to retrolens for Sagittarius A$^*$ and Messier 87$^*$ are given as $M\approx4.1\times10^6M_\odot$, $D_{OL}\approx8\times10^{-3}$ Mpc (see Table \ref{SMBH-obda}) and $M\approx6.5\times10^9M_\odot$, $D_{OL}\approx16.8$ Mpc \cite{ETH2019}, respectively.
From this figure, we find that the relative difference $\Delta(\theta_+-\theta_-)$ grows linearly with the increasing of the parameter $\alpha$. The relative difference is of order $10^{-2}$ $\mu$as for $\alpha\sim\mathcal{O}(10^{-4})$.

\begin{figure}[t]
 \centering
\begin{tabular}{cc}
\includegraphics[width=0.5 \textwidth]{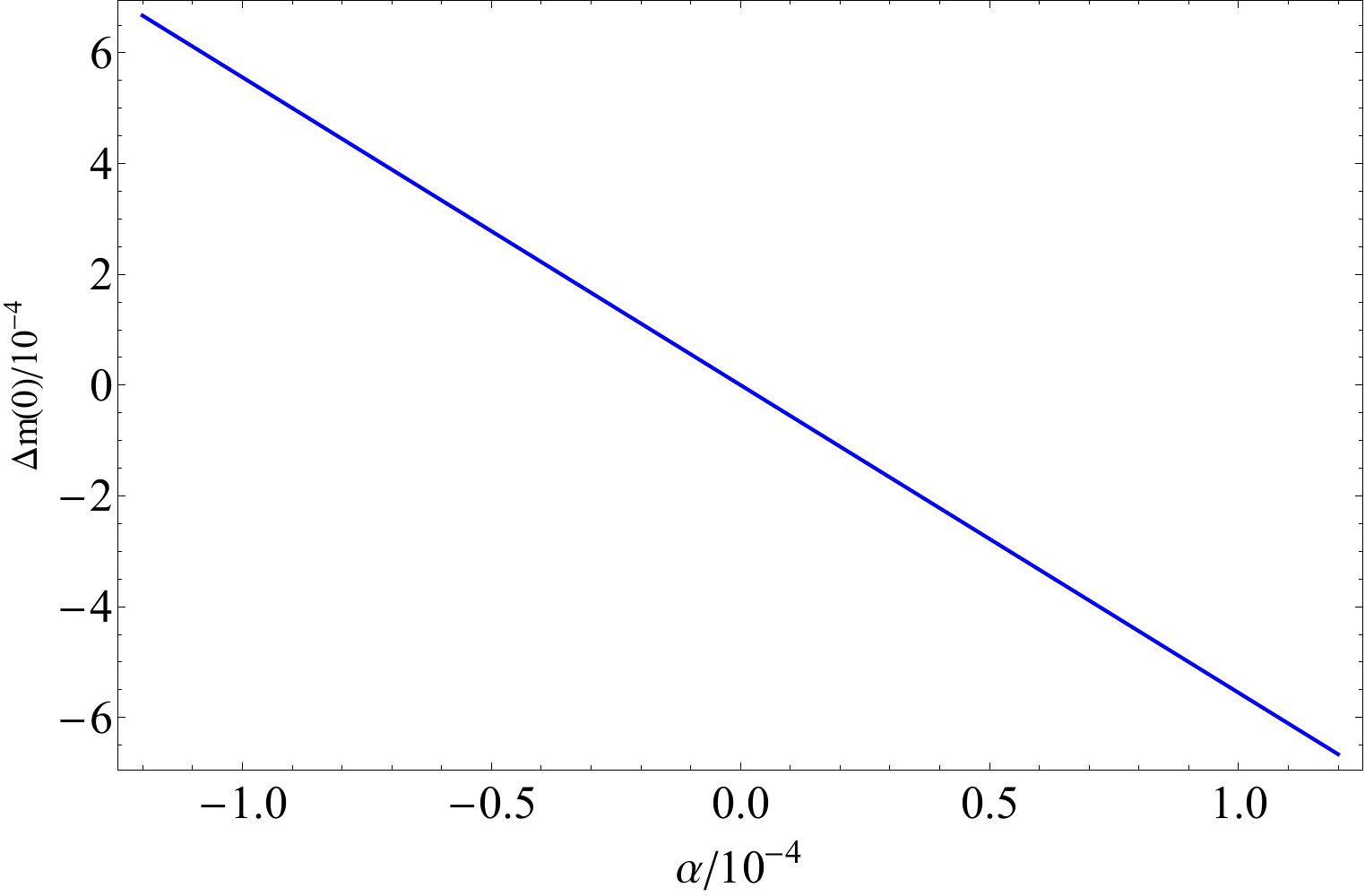}
\end{tabular}
 \caption{The relative difference $\Delta m(0)$ as a function of $\alpha$.}\label{Delmzero-MB}
\end{figure}

The magnification of the images determined by the image angles $\theta_\pm$ is given by \cite{Bozza2004}
\begin{eqnarray}
\mu_-(\beta)\simeq-\mu_+(\beta)&=&\frac{D^2_{OS}}{D^2_{LS}}s(\beta)\theta_+\frac{d\theta_+}{d\beta}\nonumber\\
&\simeq&\frac{D^2_{OS}}{D^2_{LS}}\frac{b^2_m}{D^2_{OL}}\frac{e^{\frac{\bar{b}-(2n+1)\pi}{\bar{a}}}}{\bar{a}}\left[1+e^{\frac{\bar{b}-(2n+1)\pi}{\bar{a}}}\right]s(\beta),
\end{eqnarray}
where $s(\beta)=1/\beta$ is for the point source and for the fine-size extension of source $s(\beta)$ is given by
\begin{eqnarray}
s(\beta)=\frac{D^2_{LS}}{\pi R^2_S}\int_{D_S}d\beta' d\phi,\label{Int-Sundisk}
\end{eqnarray}
where $R_S$ is the radius of the Sun, $\beta'$ and $\phi$ are the reduced radial coordinate (rescaled by $1/D_{LS}$) and azimuthal coordinate on the source plane, respectively, and $D_S$ is a uniform-luminous disk describing the Sun on the source plane \cite{Witt1994,Nemiroff1994,Alcock1997}. In general, the integral (\ref{Int-Sundisk}) can be expressed in the elliptic integrals of first and second kind \cite{Bozza2004} or can be simplified by considering the intersection point between the $\beta=0$ line (corresponding to the case of perfect alignment) and the source plane as the origin of coordinates \cite{Tsukamoto2017a,Tsukamoto2017b}. The total magnification of retrolensing double image is found as
\begin{eqnarray}
\mu_{\text{tot}}(\beta)&=&\sum_{n=0}^{\infty}\left[|\mu_+(\beta)|+|\mu_-(\beta)|\right]\nonumber\\
&=&2\frac{D^2_{OS}}{D^2_{LS}}\frac{b^2_m}{D^2_{OL}}\frac{e^{\frac{\bar{b}}{\bar{a}}}}{\bar{a}}\left[\frac{1}{2\sinh(\pi/\bar{a})}+\frac{e^{\frac{\bar{b}}{\bar{a}}}}{2\sinh(2\pi/\bar{a})}\right]|s(\beta)|.
\end{eqnarray}
The maximal magnification $\mu_{\text{tot}}(0)$ corresponds to the situation that the system of the Sun, observer, and retrolens is perfectly aligned with $s(0)$ given by $s(0)=2D_{LS}/R_S$. The apparent magnitude is defined as \cite{Geralico2003}
\begin{eqnarray}
m(\beta)=m_\odot-2.5\log_{10}\mu_{\text{tot}}(\beta),
\end{eqnarray}
where $m_\odot=-26.8$ is the optical visual magnitude of the solar luminosity. The maximal apparent magnitude is determined by $m(0)$ corresponding to the maximal brightness of the image. In Fig. \ref{Delmzero-MB}, we show the relative difference $\Delta m(0)$ between the maximal apparent magnitude predicted by our scenario and that predicted by GR.
Note that, $\Delta m(0)$ is almost independent on the retrolens mass, $D_{OL}$, $D_{OS}$, and $D_{LS}$ due to the presence of logarithm function and a fact that $\bar{a}$ and $\bar{b}$ are dependent only on the parameter $\alpha$ at the first order of their expansion in $\alpha$. We observe here that the relative difference $\Delta m(0)$ decreases monotonically as the parameter $\alpha$ increases. This suggests that compared to the Schwarzschild retrolens the maximal brightness of the image is highly amplified for the region of negative $\alpha$ parameter.
\section{\label{conclu} Conclusion}

It is possible that spacetime fundamentally has more than four dimensions and possesses hidden topological/geometric structures. In this paper, we have proposed five dimensional spacetime with a fiber fabric which is generally nontrivial. This non-trivially topological structure of spacetime allows us to resolve the hierarchy problem with the inverse size of the fifth dimension and the fundamental Planck scale to be of the order of the TeV scale. The large hierarchy between the observed Planck scale $\sim{O}(10^{18})$ GeV and the electroweak scale $\sim{O}(1)$ TeV is due to the function of $\cosh\left(\frac{\sqrt{10}}{3}\kappa_1\pi\right)$ where $\kappa_1$ is the ratio of the root-square modulus of the bulk cosmological constant to the inverse size of the fifth dimension. Here, this $\cosh$ function is physically determined from the dynamics of the horizontal metric in the vacuum of non-zero energy. The large hierarchy could be resolved with $\kappa_1\sim\mathcal{O}(35)$. In addition, the fiber fabric of spacetime allows us to resolve the problem of the chirality fermions in an elegant and natural way, which usually appears in the higher dimension theories which are the generalization of General Relativity (GR) to higher dimensions.

We have pointed out that the potential of the radion field in the present work exists at the tree level and respects to the higher dimensional general covariance. Therefore, the size of the fifth dimension or the radion mass is physically determined. Interestingly, the radion field could be responsible for a natural candidate of the inflaton. The effective action for the four dimensional metric and the radion field obtained from the fundamental action in five dimensions is naturally given in the so-called Jordan frame. In the so-called Einstein frame obtained via a properly conformal transformation, the potential for the redefined scalar field of the canonical kinetic term belongs the slow-roll inflation model of E-model. We have calculated the main observed quantities of the inflation which are the spectral index, the tensor-to-scalar ratio, the running index, and the scalar perturbation amplitude $A_s$. With the number of e-foldings about $\mathcal{O}(60)$, the inflationary observables predicted by our model are consistent with the experimental observations. Furthermore, our model predicts a sufficiently large tensor-to-scalar ratio which is around $\sim0.011$ and hence it can be tested in the near future.

Finally, we have obtained the geodesic equation which describes the motion of neutral test particles in spacetime with the fiber fabric and tested it by investigating the geodesic motion of massless and massive particles around the central gravitational bodies. We have calculated the radius of the photon sphere, the radius of the innermost stable circular orbit, the oscillating point of the near-circular orbits, the perihelion shift, the light bending angle in the weak field limit, and the observations of the strong gravitational lensing and retrolensing (another situation of strong gravitational lensing). These calculations exhibit the important differences between GR and our spacetime about the qualitative and quantitative features and therefore could allow us to distinguish it from other extensions of GR. We have compared the computed results with the observations to determine the constraints on the fiber fabric of spacetime. 

\section*{Acknowledgments}
We would like to express sincere gratitude to two referees for their constructive comments, suggestions and questions by which the quality of the paper has been improved. This research is funded by Vietnam National Foundation for Science and Technology Development (NAFOSTED) under grant number 103.01-2019.353.
\section*{Appendix}

The coefficients of the Christoffel connection given in Eq. (\ref{Chris-conn}) are explicitly expanded as
\begin{eqnarray}
\Gamma^\rho_{\mu\nu}&=&\frac{g^{\rho\lambda}}{2}(\hat{\partial}_\mu g_{\lambda\nu}+\hat{\partial}_\nu g_{\lambda\mu}-\hat{\partial}_\lambda g_{\mu\nu}).\nonumber\\
\Gamma^\theta_{\mu\nu}&=&-\frac{G^{\theta\theta}}{2}\partial_\theta g_{\mu\nu}+\frac{C^\theta_{\mu\nu}}{2}\nonumber\\
&=&\frac{\Lambda^2}{2T^2}\partial_\theta g_{\mu\nu}-\frac{g_{_X}}{2}X_{\mu\nu},\nonumber\\
\Gamma^\theta_{\mu\theta}=\Gamma^\theta_{\theta\mu}&=&\frac{G^{\theta\theta}}{2}\hat{\partial}_\mu G_{\theta\theta}\nonumber\\
&=&-\frac{\hat{\partial}_\mu T}{T},\nonumber\\
\Gamma^\nu_{\mu\theta}=\Gamma^\nu_{\theta\mu}&=&\frac{g^{\nu\lambda}}{2}\partial_\theta g_{\lambda\mu}+\frac{g_{_X}}{2\Lambda^2}T^2g^{\nu\lambda}X_{\lambda\mu},\nonumber\\
\Gamma^\mu_{\theta\theta}&=&-\frac{g^{\mu\nu}}{2}\hat{\partial}_\nu G_{\theta\theta}\nonumber\\
&=&\frac{g^{\mu\nu}}{\Lambda^2}T\hat{\partial}_\nu T,\nonumber\\
\Gamma^\theta_{\theta\theta}&=&\frac{G^{\theta\theta}}{2}\partial_\theta G_{\theta\theta}\nonumber\\
&=&-\frac{\partial_\theta T}{T}
\end{eqnarray}

The scalar curvature $\mathcal{R}^{(5)}$ of spacetime $\mathcal{M}_5$ is explicitly given as
\begin{eqnarray}
\mathcal{R}^{(5)}=G^{MN}\mathcal{R}^P_{MPN}=g^{\mu\nu}\mathcal{R}^P_{\mu P\nu}+G^{\theta\theta}\mathcal{R}^P_{\theta P\theta},
\end{eqnarray}
where
\begin{eqnarray}
g^{\mu\nu}\mathcal{R}^P_{\mu P\nu}&=&g^{\mu\nu}\left(\hat{\partial}_\nu\Gamma^M_{M \mu}-\hat{\partial}_M\Gamma^M_{\nu\mu}+\Gamma^N_{M\mu}\Gamma^M_{\nu N}-\Gamma^N_{\nu\mu}\Gamma^M_{MN}+C^N_{M\nu}\Gamma^M_{N\mu}\right)\nonumber\\
&=&g^{\mu\nu}\left(\hat{\partial}_\nu\Gamma^\lambda_{\lambda\mu}-\hat{\partial}_\lambda\Gamma^\lambda_{\nu\mu}+\Gamma^\rho_{\lambda\mu}\Gamma^\lambda_{\nu\rho}-\Gamma^\rho_{\nu\mu}\Gamma^\lambda_{\lambda\rho}\right)+g^{\mu\nu}\left(\hat{\partial}_\nu\Gamma^\theta_{\theta\mu}+\Gamma^\rho_{\theta\mu}\Gamma^\theta_{\nu\rho}+\Gamma^\theta_{\lambda\mu}\Gamma^\lambda_{\nu\theta}+\Gamma^\theta_{\theta\mu}\Gamma^\theta_{\nu\theta}\right)\nonumber\\
&&-g^{\mu\nu}\left(\partial_\theta\Gamma^\theta_{\nu\mu}-\Gamma^\theta_{\nu\mu}\Gamma^\lambda_{\lambda\theta}-\Gamma^\rho_{\nu\mu}\Gamma^\theta_{\theta\rho}-\Gamma^\theta_{\nu\mu}\Gamma^\theta_{\theta\theta}\right)+g^{\mu\nu}C^\theta_{\lambda\nu}\Gamma^\lambda_{\theta\mu},                                                                                                                                                                                                                                                                                                                                                                                                                                                                                                                                                                                                   \label{sccurI}\\
G^{\theta\theta}\mathcal{R}^P_{\theta P\theta}&=&G^{\theta\theta}\left(\partial_\theta\Gamma^M_{M\theta}-\partial_M\Gamma^M_{\theta\theta}+\Gamma^N_{M\theta}\Gamma^M_{\theta N}-\Gamma^N_{\theta\theta}\Gamma^M_{MN}+C^N_{M\theta}\Gamma^M_{N\theta}\right)\nonumber\\
&=&G^{\theta\theta}\left(\partial_\theta\Gamma^\mu_{\mu\theta}-\hat{\partial}_\mu\Gamma^\mu_{\theta\theta}+\Gamma^\nu_{\mu\theta}\Gamma^\mu_{\theta\nu}+\Gamma^\mu_{\theta\theta}\Gamma^\theta_{\theta\mu}-\Gamma^\mu_{\theta\theta}\Gamma^\nu_{\nu\mu}-\Gamma^\theta_{\theta\theta}\Gamma^\mu_{\mu\theta}\right).\label{sccurII}      
\end{eqnarray}
In (\ref{sccurI}) and (\ref{sccurII}),  we find the following combinations
\begin{eqnarray}
g^{\mu\nu}\left(\hat{\partial}_\nu\Gamma^\theta_{\theta\mu}-\Gamma^\rho_{\nu\mu}\Gamma^\theta_{\theta\rho}\right)-G^{\theta\theta}\Gamma^\mu_{\theta\theta}\Gamma^\theta_{\theta\mu}&\equiv&\nabla_MY^M_1,\nonumber\\
g^{\mu\nu}\left(\partial_\theta\Gamma^\theta_{\nu\mu}+\Gamma^\theta_{\nu\mu}\Gamma^\lambda_{\lambda\theta}+\Gamma^\theta_{\nu\mu}\Gamma^\theta_{\theta\theta}\right)&\equiv&\nabla_MY^M_2+\frac{G^{\theta\theta}}{2}\partial_\theta g^{\mu\nu}\partial_\theta g_{\mu\nu},\nonumber\\
G^{\theta\theta}\left(\partial_\theta\Gamma^\mu_{\mu\theta}-\Gamma^\theta_{\theta\theta}\Gamma^\mu_{\mu\theta}\right)-g^{\mu\nu}\Gamma^\theta_{\mu\nu}\Gamma^\lambda_{\lambda\theta}&\equiv&\nabla_MY^M_3,\nonumber\\
G^{\theta\theta}\left(\hat{\partial}_\mu\Gamma^\mu_{\theta\theta}+\Gamma^\mu_{\theta\theta}\Gamma^\nu_{\nu\mu}+\Gamma^\mu_{\theta\theta}\Gamma^\theta_{\theta\mu}\right)&\equiv&\nabla_MY^M_4+\frac{g^{\mu\nu}}{2}\hat{\partial}_\mu G^{\theta\theta}\hat{\partial}_\nu G_{\theta\theta},
\end{eqnarray}
where
\begin{eqnarray}
Y^M_1&\equiv&\left(g^{\mu\nu}\Gamma^\theta_{\theta\mu},0\right),\nonumber\\
Y^M_2&\equiv&\left(0,g^{\mu\nu}\Gamma^\theta_{\nu\mu}\right),\nonumber\\
Y^M_3&\equiv&\left(0,G^{\theta\theta}\Gamma^\mu_{\mu\theta}\right),\nonumber\\
Y^M_4&\equiv&\left(G^{\theta\theta}\Gamma^\mu_{\theta\theta},0\right).
\end{eqnarray}
Note that, the vectors $Y^M_1$ and $Y^M_4$ have the zero vertical components, whereas the vectors $Y^M_2$ and $Y^M_3$ have the zero horizontal components. The terms $\nabla_MY^M_i$ with $i=1,2,3,4$ are divergences which correspond to the boundary terms and hence vanish at infinity.


\begin{thebibliography}{99}
\bibitem{Kaluza1921} T. Kaluza, Sitzungsber. Preuss. Akad. Wiss. Berlin
(Math. Phys. ) {\bf 1921}, 966 (1921).

\bibitem{Klein1926} O. Klein, Z. Phys. {\bf 37}, 895 (1926).

\bibitem{Dienes1998} K. R. Dienes, E. Dudas, and T. Gherghetta, Phys. Lett. B {\bf 436}, 55 (1998).

\bibitem{Antoniadis1990} I. Antoniadis, Phys. Lett. B {\bf 246}, 377 (1990).

\bibitem{Arkani-Hamed2000} N. Arkani-Hamed and M. Schmaltz, Phys. Rev. D {\bf 61}, 033005 (2000).

\bibitem{Arkani-Hamed1998} N. Arkani-Hamed, S. Dimopoulos, and G. R. Dvali, Phys. Lett. B {\bf 429}, 263 (1998).

\bibitem{Antoniadis1998} I. Antoniadis, N. Arkani-Hamed, S. Dimopoulos, and G. R. Dvali, Phys. Lett. B {\bf 436},
257 (1998).

\bibitem{Randall1999} L. Randall and R. Sundrum, Phys. Rev. Lett. {\bf 83}, 3370 (1999).

\bibitem{Kaloper2000} N. Kaloper, J. March-Russell, G. D. Starkman, and M. Trodden, Phys. Rev. Lett. {\bf 85}, 928
(2000).

\bibitem{Nath1999} P. Nath and M. Yamaguchi, Phys. Rev. D {\bf 60}, 116004 (1999).

\bibitem{Masip1999} M. Masip and A. Pomarol, Phys. Rev. D {\bf 60}, 096005 (1999).

\bibitem{Casalbuoni1999} R. Casalbuoni, S. De Curtis, D. Dominici, and R. Gatto, Phys. Lett. B {\bf 462}, 48 (1999).

\bibitem{Carone2000} C. D. Carone, Phys. Rev. D {\bf 61}, 015008 (2000).

\bibitem{Appelquist2001} T. Appelquist, H.-C. Cheng, and B. A. Dobrescu, Phys. Rev. D {\bf 64}, 035002 (2001). 

\bibitem{Csaki2004} C. Csaki, J. Hubisz, and P. Meade, ``TASI lectures on electroweak symmetry breaking from extra dimensions", arXiv: hep-ph/0510275.

\bibitem{Starobinsky1980} A. A. Starobinsky, Phys. Lett. B {\bf 91}, 99 (1980).

\bibitem{Guth1981} A. H. Guth, Phys. Rev. D {\bf 23}, 347 (1981).

\bibitem{Linde1983} A. D. Linde, Phys. Lett. B {\bf 108}, 389 (1983).

\bibitem{Albrecht1982} A. Albrecht and P. J. Steinhardt, Phys. Rev. Lett. {\bf 48}, 1220 (1982).



\bibitem{Cline2000} J. M. Cline, Phys. Rev. D {\bf 61}, 023513 (2000).

\bibitem{Kolb2003} E. W. Kolb, G. Servant, and T. M. P. Tait, JCAP {\bf 0307}, 008 (2003).

\bibitem{Sundrum2010} R. Sundrum and C. M. Wells, JHEP {\bf 02}, 097 (2010).

\bibitem{Trudeau2012} J. Trudeau and J. M. Cline, JHEP {\bf 02}, 081 (2012).

\bibitem{Fukazawa2013} Y. Fukazawa, T. Inami, and Y. Koyama, Prog. Theor. Exp. Phys. {\bf 2013}, 021B01 (2013).

\bibitem{Bahamonde2019} S. Bahamonde, K. Flathmann, and C. Pfeifer, Phys. Rev. D {\bf 100}, 084064 (2019).

\bibitem{GuoMiao2020} Y. Guo and Y.-G. Miao, Phys. Rev. D {\bf 102}, 084057 (2020).

\bibitem{Boonserm2020} P. Boonserm, T. Ngampitipan, A. Simpson, and M. Visse, Phys. Rev. D {\bf 101}, 024050 (2020)

\bibitem{Turimov2020} B. Turimov, J. Rayimbaev, A. Abdujabbarov, B. Ahmedov, and Z. Stuchl\'{i}k, Phys. Rev. D {\bf 102}, 064052 (2020).

\bibitem{XXZeng2020} X.-X. Zeng, H.-Q. Zhang, and H. Zhang, Eur. Phys. J. C {\bf 80}, 872 (2020).

\bibitem{GuoLi2020} M. Guo and P.-C. Li, Eur. Phys. J. C {\bf 80}, 588 (2020).


\bibitem{Salam1982} A. Salam and J. Strathdee, Annals of Physics {\bf 141}, 316 (1982).


\bibitem{Nakahara2003} M. Nakahara, \emph{Geometry, Topology and Physics}, (Institute of
Physics Publishing, London, 2003).

\bibitem{Tanabashi2018} M. Tanabashi \textit{et al.} (Particle Data Group), Phys. Rev. D {\bf 98}, 030001 (2018).

\bibitem{Nam2019a} C. H. Nam, Eur. Phys. J. C {\bf 79}, 384 (2019).

\bibitem{Nam2020a} C. H. Nam, Eur. Phys. J. C {\bf 80}, 231 (2020).

\bibitem{Nam2020b} C. H. Nam, arXiv: 2001.02421 [hep-ph].

\bibitem{Planck2018} Y. Akrami \textit{et al.} (Planck Collaboration), Astron. Astrophys. {\bf 641}, A10 (2020).

\bibitem{Kerr1963} R. P. Kerr, Phys. Rev. Lett. {\bf 11}, 237 (1963).

\bibitem{Beck2000} J. G. Beck, Solar Phys. {\bf 191}, 47 (2000).

\bibitem{Fuller2019} J. Fuller and L. Ma, Astrophys. J. Lett. {\bf 881}, L1 (2019).

\bibitem{Fragos2015} T. Fragos and J. E. McClintock, Astrophys. J. {\bf 800}, 17 (2015). 

\bibitem{Miler2014} M. C. Miller and J. M. Miller, Phys. Rept. {\bf 548}, (2014) 1-34.

\bibitem{Berti2008} E. Berti and M. Volonteri, Astrophys. J. {\bf 684}, 822 (2008).

\bibitem{Roulet2018} J. Roulet and M. Zaldarriaga, Mon. Not. Roy. Astron. Soc. {\bf 484}, 4216 (2019).

\bibitem{LIGO2018} The LIGO Scientific Collaboration, Astrophys. J. Lett. {\bf 882}, L24 (2019).

\bibitem{CMWill2006} C. M. Will, Living Rev. Rel. {\bf 9}, 3 (2006).


\bibitem{GRbook-Schutz} B. Schutz, \textit{A First Course in General Relativity}, Cambridge University Press, Cambridge, UK, 2009.

\bibitem{Shapiro2004} S. Shapiro, J. Davis, D. Lebach, and J. Gregory, Phys. Rev. Lett. {\bf 92}, 121101 (2004).

\bibitem{Will2014} C. M. Will, Living Rev. Rel. {\bf 17}, 4 (2014).

\bibitem{Robertson1991} D. S. Robertson, W. E. Carter, and W. Dillinger, Nature {\bf 349}, 768 (1991).


\bibitem{Ghez2008} A. M. Ghez \emph{et al.}, Astrophys. J. {\bf 689}, 1044 (2008).

\bibitem{Kafle2018} P. R. Kafle, S. Sharma, G. F. Lewis, A. S. G. Robotham, and
S. P. Driver, Mon. Not. R. Astron. Soc. {\bf 475}, 4043 (2018).

\bibitem{Murphy2011} J. D. Murphy, K. Gebhardt, and J. J. Adams, Astrophys. J. {\bf 729}, 129 (2011).

\bibitem{Virbhadra1998} K. S. Virbhadra, D. Narasimha, and S. M. Chitre, Astron. Astrophys. {\bf 337} (1998) 1.

\bibitem{Bozza2002} V. Bozza, Phys. Rev. D {\bf 66}, 103001 (2002).

\bibitem{Shapiro1964} I. I. Shapiro, Phys. Rev. Lett. {\bf 13}, 789 (1964).

\bibitem{Bertotti2003} B. Bertotti, L. Iess, and P. Tortora, Nature {\bf 425}, 374
(2003).



\bibitem{Ohanian1987} H. C. Ohanian, Am. J. Phys. {\bf 55}, 428 (1987).

\bibitem{Bozza2004} V. Bozza and L. Mancini, Astrophys. J. {\bf 611}, 1045 (2004).

\bibitem{ETH2019} The Event Horizon Telescope Collaboration, Astrophys. J. Lett. {\bf 875}, L6 (2019).

\bibitem{Witt1994} H. J. Witt and S. Mao, Astrophys. J. {\bf 430}, 505 (1994).

\bibitem{Nemiroff1994} R. J. Nemiroff and W. A. D. T. Wickramasinghe, Astrophys. J. {\bf 424}, L21 (1994).

\bibitem{Alcock1997} C. Alcock \emph{et al.} [MACHO and GMAN Collaborations], Astrophys. J. {\bf 491}, 436 (1997).

\bibitem{Tsukamoto2017a} N. Tsukamoto and Y. Gong, Phys. Rev. D {\bf 95}, 064034 (2017).

\bibitem{Tsukamoto2017b} N. Tsukamoto, Phys. Rev. D {\bf 95}, 084021 (2017).

\bibitem{Geralico2003} F. D. Paolis, A. Geralico, G. Ingrosso, and A. A. Nucita, Astron. Astrophys. {\bf 409}, 809 (2003).
\end{thebibliography}
\end{document}